%% file: 0PaperID2079.tex
\documentclass[sigconf]{acmart}
\usepackage{booktabs}      
\usepackage{multirow}      
\usepackage{array}         
\usepackage{float}         
\usepackage[normalem]{ulem}

\usepackage{amsmath,amsfonts,amssymb} 
\usepackage{mathrsfs}      
\usepackage{bm}
\usepackage{algorithm}     
\usepackage{algpseudocode} 
\floatname{algorithm}{Algorithm} 
\usepackage{graphicx}      
\usepackage{subcaption}    
\usepackage{enumitem}      
\usepackage{multicol}      
\usepackage{comment}      
\usepackage{footmisc}      
\usepackage{xcolor}       
\usepackage{xspace}       
\usepackage{hyperref}  
\hypersetup{
  colorlinks=true,
  pdfborder={0 0 0}
}

\setcopyright{acmlicensed}
\copyrightyear{2026}
\acmYear{2026}
\setcopyright{cc}
\setcctype{by-nc-nd}
\acmConference[WWW '26]{Proceedings of the ACM Web Conference 2026}{April 13--17, 2026}{Dubai, United Arab Emirates}
\acmBooktitle{Proceedings of the ACM Web Conference 2026 (WWW '26), April 13--17, 2026, Dubai, United Arab Emirates}
\acmPrice{}
\acmISBN{979-8-4007-2307-0/2026/04}
\acmDOI{10.1145/3774904.3792408}
\begin{document}
\title{BlossomRec: Block-level Fused Sparse Attention Mechanism for Sequential Recommendations}


\author{Mengyang Ma}
\email{ronin.ma@my.cityu.edu.hk}
\authornote{Authors contribute equally to this research.}
\orcid{0009-0007-0426-6008}
\affiliation{%
  \institution{City University of Hong Kong}
  \city{Hong Kong}
  \country{China}}

\author{Xiaopeng Li}
\email{xiaopli2-c@my.cityu.edu.hk}
\authornotemark[1]
\affiliation{%
  \institution{City University of Hong Kong}
  \city{Hong Kong}
  \country{China}
}

\author{Wanyu Wang}
\email{wanyuwang4-c@my.cityu.edu.hk}
\authornotemark[1]
\affiliation{%
  \institution{City University of Hong Kong}
  \city{Hong Kong}
  \country{China}
}

\author{Zhaocheng Du}
\email{duzhaocheng1998@gmail.com}
\affiliation{%
 \institution{Independent Researcher}
 \country{China}}

\author{Jingtong Gao}
\email{jt.g@my.cityu.edu.hk}
\affiliation{%
  \institution{City University of Hong Kong}
  \city{Hong Kong}
  \country{China}}

\author{Pengyue Jia}
\email{jia.pengyue@my.cityu.edu.hk}
\affiliation{%
  \institution{City University of Hong Kong}
  \city{Hong Kong}
  \country{China}}

\author{Yuyang Ye}
\email{yuyang.ye@rutgers.edu}
\affiliation{%
  \institution{Rutgers University}
  \city{}
  \state{New Jersey}
  \country{United States}}

\author{Yiqi Wang}
\email{wangy206@msu.edu}
\affiliation{%
  \institution{Michigan State University}
  \city{}
  \state{Michigan}
  \country{United States}}

\author{Yunpeng Weng}
\email{wengyp@mail3.sysu.edu.cn}
\affiliation{%
  \institution{Tencent}
  \city{Shenzhen}
  \country{China}}

\author{Weihong Luo}
\email{lobby66@163.com}
\affiliation{%
  \institution{Tencent}
  \city{Shenzhen}
  \country{China}}

\author{Xiao Han}
\authornote{Corresponding author.}

\email{hahahenha@gmail.com}
\affiliation{%
  \institution{Zhejiang University of Technology}
  \city{Hangzhou}
  \country{China}}

\author{Xiangyu Zhao}
\email{xianzhao@cityu.edu.hk}
\affiliation{%
  \institution{City University of Hong Kong}
  \city{Hong Kong}
  \country{China}}

\renewcommand{\shortauthors}{Mengyang Ma et al.}

\begin{abstract}
  Transformer architectures have been widely used in sequential recommender systems (SRS). However, as user interaction histories grow longer, the computational cost and memory consumption of these models increase substantially. This is mainly caused by the standard attention mechanism. Although there exist many methods employing efficient attention and SSM-based models, these approaches struggle to effectively model long sequences and may exhibit unstable performance on short sequences. To address these challenges, we design a sparse attention mechanism, BlossomRec, which models both long-term and short-term user interests through attention computation to achieve stable performance across sequences of varying lengths. Specifically, we categorize user interests in recommendation systems into long-term and short-term interests, and compute them using two distinct sparse attention patterns, with the results combined through a learnable gated output. Theoretically, it significantly reduces the number of interactions participating in attention computation. Extensive experiments on four public datasets demonstrate that BlossomRec, when integrated with state-of-the-art Transformer-based models, achieves comparable or even superior performance while significantly reducing memory usage, providing strong evidence of BlossomRec's efficiency and effectiveness. The code is available at \url{https://github.com/Applied-Machine-Learning-Lab/WWW2026_BlossomRec}.
  
\end{abstract}



\maketitle
\input{1Introduction}

\input{2PRELIMINARY}

\input{3Framework}

\input{4Experiment}

\input{5Relatedwork}

\input{6Conclusion}
\bibliographystyle{ACM-Reference-Format}
\balance
\bibliography{9References}

\appendix
\input{10Appendix}

\end{document}

%% file: 1Introduction.tex
\section{Introduction}
Sequential recommender systems (SRS) have been widely applied in streaming media \cite{covington2016youtube,lu2025liveforesighter,zhao2025joint}, e-commerce \cite{chai2025longer, zhai2024actions}, and social media \cite{huang2018csan,feng2024long, li2025survey} in recent years. With the advancement of machine learning, various neural network models have been employed \cite{kang2018self,jannach2017recurrent,li2023hamur,li2023agent4ranking,jia2024d3,gao2024hierrec,gao2025samplellm,li2025survey}, among which transformer-based models have achieved remarkable performance \cite{sun2019bert4rec,kang2018self,li2020time,du2022contrastive}.
However, as user interaction histories routinely exceed thousands of entries, standard attention-based transformer models face quadratic complexity when processing such long sequences and may fail to adequately capture users' long-term interests. 
Recent LLM-based recommenders~\cite{chen2024hllm,liu2024llm,liu2025bridge} further highlight the importance of modeling long contexts, but they often introduce additional inference latency and computational overhead for long user sequences~\cite{cui2024distillation,bao2025bi,geng2024breaking}.
How to effectively model both long-term and short-term user interests from long sequences under strict computational resource constraints has become a critical challenge.

Existing studies have primarily addressed the efficiency of long-sequence SRSs along three lines of investigation. Linear-attention approaches, such as LinRec~\cite{liu2023linrec}, approximate dense attention to reduce computational cost. Efficient transformer models, such as STRec~\cite{li2023strec}, exploit the sparsity characteristics of SRSs to reduce memory overhead and inference time. State space model (SSM)-based methods replace attention with recurrent state-space updates~\cite{liu2024mamba4rec,zhang2025m2rec}; for example, Mamba4Rec~\cite{liu2024mamba4rec} adapts Mamba to sequential recommendations and achieves linear complexity with hardware-aware computation. These methods substantially improve the efficiency of long-sequence recommendations.

Despite their progress, three limitations remain. First, models based on efficient transformers~\cite{li2023strec} and linear attention~\cite{liu2023linrec} tend to over-emphasize recent interactions. Although this approach has proven effective, the model may fail to capture users' long-term interests when confronted with long sequences, leading to performance degradation. Second, SSM-based models~\cite{liu2024mamba4rec} may struggle to effectively model both long and short sequences, resulting in insufficient stability of their results~\cite{zhang2025glint}. Third, SSM-based models, owing to architectural modifications, exhibit increased deployment complexity~\cite{gu2023mamba} and lack compatibility with existing models.

To address these limitations, we propose BlossomRec, a \textbf{blo}ck-level fused \textbf{s}par\textbf{s}e attenti\textbf{o}n \textbf{m}echanism for sequential recommendation. First, we integrate long-term and short-term interest modeling in SRS~\cite{yu2026malloc, lv2019sdm,shen2022hierarchically,zhang2025glint} into the attention mechanism. By employing two types of sparse attention computation to model long-term and short-term interests separately, our approach can effectively model user interaction histories across long and short sequences while maintaining stable performance. Second, through empirical observations (Appendix~\ref{sec:sparse}) of real-world user sequences, we adopt block-level modeling. For long-term interest modeling, we selectively compute attention by calculating attention scores for chunked sequences. This selective attention mechanism significantly reduces the number of interactions required for computation, thereby improving efficiency. For short-term interest modeling, we employ a power-law-based sparse attention mask~\cite{chen2025powerattention,li2019logsparse} that reduces computational costs while preserving the receptive field. Finally, we introduce a learnable gating module to adaptively fuse the two attention outputs, enhancing result stability. Third, our attention mechanism is theoretically compatible with various transformer-based models, facilitating ease of deployment. The major contributions of this work are summarized as follows:
\begin{itemize}[leftmargin=*]
    \item Building upon prior research \cite{li2023strec, jannach2017recurrent} and empirical observations of long user sequences~(Appendix~\ref{sec:sparse}), we show that partitioning sequences into blocks provides an effective inductive bias for capturing interest dynamics, and that selecting a sparse subset of blocks is sufficient to model long-term user interests accurately. 
    \item We propose a novel block-level fused sparse attention mechanism that dynamically models long-term and short-term interests through two complementary sparse pathways—using importance-based block selection for long-range dependencies and recency-aware masking for short-term contexts—with a learnable gating fusion strategy. 
    \item Experiments on four public benchmark datasets demonstrate that our model achieves performance comparable to or even surpassing that of state-of-the-art models while demonstrating less memory consumption and higher computational speed.
\end{itemize}

%% file: 2PRELIMINARY.tex
\section{Preliminaries}
In this section, we define the sequential recommendation task and then introduce standard, multi-head, and grouped query attention mechanisms used in our framework. 
\subsection{Definition of Sequential Recommendation Task}

In sequential recommendations, we consider a set of users $\mathcal{U} = \{u_1, u_2, \ldots, u_{|\mathcal{U}|}\}$ and a set of items $\mathcal{V} = \{v_1, v_2, \ldots, v_{|\mathcal{V}|}\}$. Each user $u_i \in \mathcal{U}$ has an ordered sequence of historical interactions denoted as $s_i = [v_1^{(i)}, v_2^{(i)}, \ldots, v_{n_i}^{(i)}]$, where $n_i$ represents the length of user $u_i$'s interaction sequence. The primary objective is to develop an efficient recommendation framework to predict the next item a user will interact with, given their historical interactions.
\subsection{Attention Mechanisms}

\textbf{Standard Self-Attention.} In Transformer-based SRS, the architecture is typically composed of an embedding layer, encoder layers, and a prediction layer, among which self-attention mechanism constitutes the core component of the encoder layer~\cite{kang2018self,sun2019bert4rec}.  Given an input sequence, the self-attention mechanism first projects it into three matrices: query ($\boldsymbol{Q}$), key ($\boldsymbol{K}$), and value ($\boldsymbol{V}$), through learned linear transformations. The attention output is computed as:
\begin{equation}
    \text{Attn}(\boldsymbol{Q},\boldsymbol{K},\boldsymbol{V},\boldsymbol{M}) = \text{softmax}\left( \frac{\boldsymbol{Q}\boldsymbol{K}^\top}{\sqrt{d_k}} + \log\boldsymbol{M} \right) \boldsymbol{V}
\end{equation}
where $\boldsymbol{M} \in \mathbb{R}^{n \times n}$ is a mask matrix. $d_k$ represents the dimension of the key vectors. The computational complexity of self-attention is $\mathcal{O}(n^2d)$, where $n$ is the sequence length and $d$ is the hidden dimension, making it computationally expensive for long sequences.

\noindent \textbf{Multi-Head Attention.} Multi-head attention~\cite{vaswani2017attention} employs multiple attention mechanisms in parallel. Each head independently learns different aspects of the input representations. The multi-head attention is defined as:
\begin{equation}
\mathrm{MultiHead}(\boldsymbol{Q},\boldsymbol{K},\boldsymbol{V})=
\mathrm{Concat}\bigl(\mathrm{head}_1,\dots,\mathrm{head}_h\bigr) \boldsymbol{W}^{O}
\end{equation}
\begin{equation}
\mathrm{head}_i=
\mathrm{Attn}(\boldsymbol{Q}\boldsymbol{W}_i^Q,\,\boldsymbol{K}\boldsymbol{W}_i^{K},\,\boldsymbol{V}\boldsymbol{W}_i^{V})
\end{equation}
where $h$ is the number of attention heads, $\boldsymbol{W}_i^Q, \boldsymbol{W}_i^K, \boldsymbol{W}_i^V$ are per-head projection matrices, and $\boldsymbol{W}^O$ is the output projection matrix.
 \begin{figure*}
	\centering
	\includegraphics[width=0.8\linewidth]{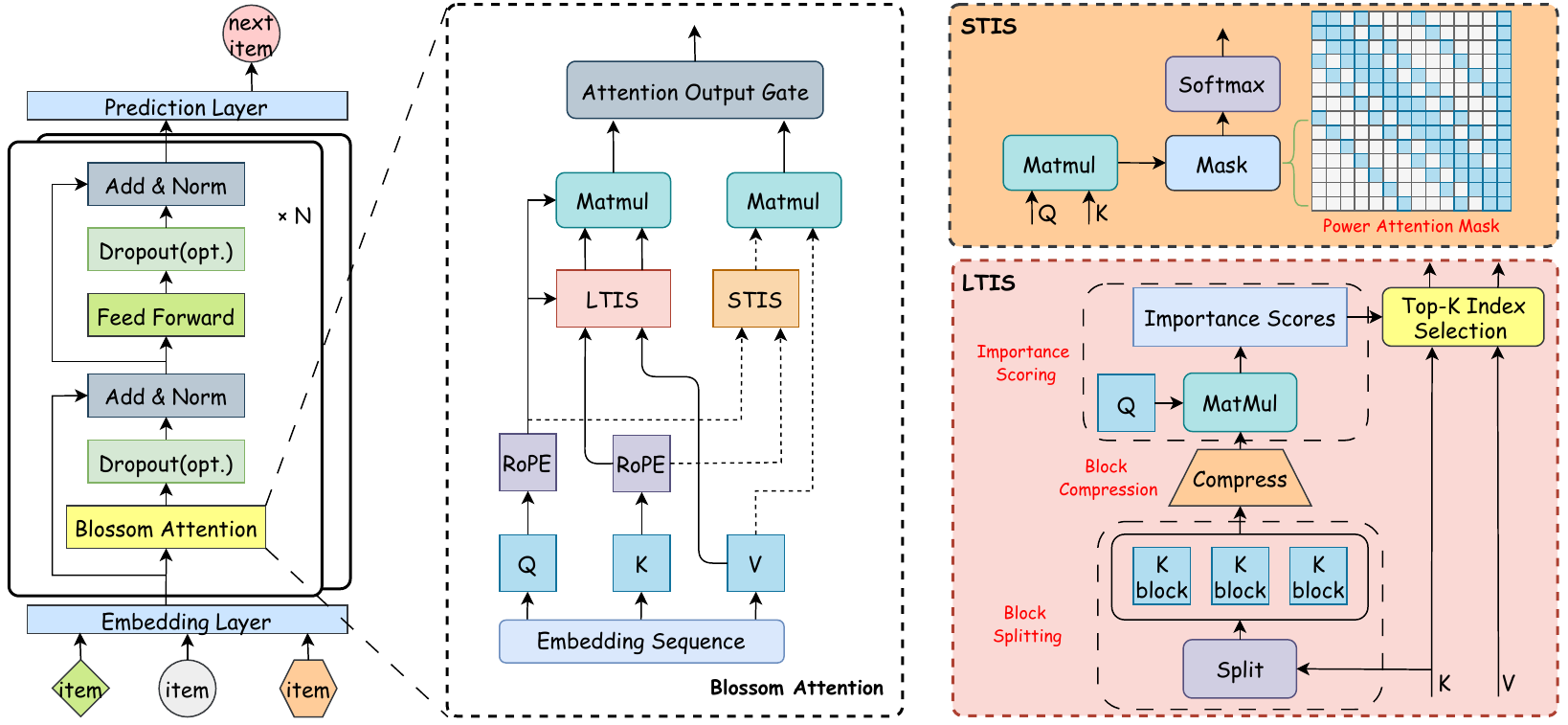}
	\caption{Overview of the BlossomRec framework.}
	\label{fig:Fig1_Overview}
\end{figure*}

\noindent \textbf{Grouped Query Attention (GQA).} GQA is used in BlossomRec to compute attention from $\boldsymbol{Q}$, $\boldsymbol{K}$, $\boldsymbol{V}$. Grouped Query Attention~\cite{ainslie2023gqa} is more efficient than multi-head attention by sharing key and value projections across multiple query heads. GQA organizes the query heads into $g$ groups, each sharing the same key and value projections. This can be formulated as:
\begin{equation}
\text{GQA}(\boldsymbol{Q}, \boldsymbol{K}, \boldsymbol{V}) = \text{Concat}(\text{head}_1, \ldots, \text{head}_h)\boldsymbol{W}^O
\end{equation}
\begin{equation}
\text{head}_i = \text{Attn}(\boldsymbol{Q}_i, \boldsymbol{K}_{g(i)},\boldsymbol{V}_{g(i)})
\end{equation}
where $h$ is the number of query heads, $g(i) = \left\lceil \frac{i g}{h} \right\rceil$ is $KV$ group index for head $i$, $g$ is the number of KV groups ($g < h$).

%% file: 3Framework.tex
\section{Framework}
In this section, we introduce the BlossomRec framework in detail.

\subsection{Overview}

BlossomRec is designed as a plug-in sparse attention mechanism for Transformer-based SRS. As illustrated in Figure~\ref{fig:Fig1_Overview}, BlossomRec replaces the standard full attention module with Blossom Attention while keeping the overall Transformer backbone unchanged. Blossom Attention contains two complementary pathways. Long-Term Interest Selection (LTIS) selects informative key--value blocks to capture long-range user preferences, while Short-Term Interest Selection (STIS) applies a structured sparse mask to preserve recent and local behavioral patterns. The two outputs are then adaptively fused through a learnable gating module, enabling the model to balance long-term and short-term interests across sequences of different lengths. This design reduces the computational cost of attention compared with standard full attention while preserving the capacity to model both global and local user interests.

\subsection{Embedding Layer}

The embedding layer maps each item in a user interaction sequence to a dense representation. Given an input sequence $s = [v_1, v_2, \ldots, v_L]$, where $L$ is the sequence length, each item $v_t$ is mapped to a $d$-dimensional embedding through a learnable item embedding table $\boldsymbol{W}_{e} \in \mathbb{R}^{|\mathcal{V}| \times d}$:
\begin{equation}
    \boldsymbol{e}_t = \boldsymbol{W}_{e}[v_t]
\end{equation}
where $\mathcal{V}$ denotes the item set and $\boldsymbol{W}_{e}[v_t]$ denotes the row of $\boldsymbol{W}_{e}$ indexed by item $v_t$. Since Rotary Position Embedding (RoPE)~\cite{su2024roformer} is applied to the query and key representations in the attention module, the embedding layer does not use explicit absolute position embeddings. The encoded sequence is denoted as:
$\boldsymbol{E} = [\boldsymbol{e}_1, \boldsymbol{e}_2, \cdots, \boldsymbol{e}_L]^T$.
\subsection{Block-Level Fused Sparse Attention Mechanism}
A key challenge in sparse attention for sequential recommendation is to improve efficiency while retaining the ability to model useful long-range and local dependencies. As observed in Figure~\ref{fig:app1}(Appendix~\ref{sec:sparse}), long user interaction sequences can often be processed effectively through block-wise operations. Guided by this observation, Blossom Attention adopts a block-level fused sparse attention strategy. As shown in Figure~\ref{fig:Fig1_Overview}, the encoded sequence is first projected into query, key, and value matrices:
\begin{equation}
    \boldsymbol{Q} = \boldsymbol{E} \cdot \boldsymbol{W}^Q, \quad 
    \boldsymbol{K} = \boldsymbol{E} \cdot \boldsymbol{W}^K, \quad 
    \boldsymbol{V} = \boldsymbol{E} \cdot \boldsymbol{W}^V
\end{equation}
where $\boldsymbol{W}^{Q}, \boldsymbol{W}^{K} \in \mathbb{R}^{d \times d_k}$ and $\boldsymbol{W}^{V} \in \mathbb{R}^{d \times d_v}$ are learnable projection matrices. RoPE~\cite{su2024roformer} is then applied to $\boldsymbol{Q}$ and $\boldsymbol{K}$ to encode relative positional information. The resulting $\boldsymbol{Q}$, $\boldsymbol{K}$, and $\boldsymbol{V}$ are fed into LTIS and STIS in parallel to model long-term and short-term user interests, respectively.

\subsection{Long-Term Interest Selection (LTIS)}
\label{sec:LTIS}
Long-term interests manifest as intrinsic user preferences that often reflect relatively stable user preferences across an interaction sequence~\cite{shen2022hierarchically}. The Long-Term Interest Selection (LTIS) module is designed to capture long-term user interests from long interaction sequences. Furthermore, as observed in Appendix~\ref{sec:sparse}, such interests can be modeled by splitting user sequences into blocks. These blocks are allowed to overlap to more faithfully approximate the distribution of long-term interests. To identify blocks that encapsulate these interests, $\boldsymbol{K}$ is first partitioned into blocks, which are then compressed. The compressed attention scores are subsequently used as the selection criterion. Introducing a stride parameter further preserves the distributional continuity of interest blocks. Leveraging compressed attention scores also reduces computational cost. After applying LTIS, only selected key--value (KV) blocks enter the attention computation, allowing the model to capture global interaction signals while attending only to a sparsified subset of interactions.

\noindent \textbf{Block Splitting.} Given $\boldsymbol{K} \in \mathbb{R}^{L \times d_k}$ derived from a user interaction sequence, LTIS partitions it into overlapping blocks:
\begin{equation}
\boldsymbol{K}_i = \boldsymbol{K}[is+1 : is+l], 
\quad i = 0, 1, \ldots, M-1
\end{equation}
where $l$ is the compression block size, $s$ is the stride, and $M=\left\lfloor (L-l)/s \right\rfloor+1$ is the number of compressed blocks. The same block partitioning is applied to $\boldsymbol{V}$.

\noindent \textbf{Block Compression.} Each key block is compressed into a representative vector through a learnable MLP $\varphi(\cdot)$:
\begin{equation}
\tilde{\boldsymbol{K}}_{\text{L}}^{cmp}
= f_{\text{K}}^{cmp}\bigl(\boldsymbol{K}_{1:L}\bigr)
= \Bigl\{\,
\varphi(\boldsymbol{K}_i)
\Bigr\}_{i=0}^{M-1}
\label{eq:block-compress}
\end{equation}
where $\tilde{\boldsymbol{K}}_{\mathrm{L}}^{\mathrm{cmp}}$ denotes the compressed key matrix. The same procedure is applied to $\boldsymbol{V}$, producing $\tilde{\boldsymbol{V}}_{\mathrm{L}}^{\mathrm{cmp}}$.

\noindent \textbf{Importance Scoring.} LTIS estimates the importance of compressed blocks by computing attention scores between query representations and compressed keys:
\begin{equation}
\boldsymbol{S}_L^{cmp} = \operatorname{softmax}\!\left(
\boldsymbol{Q}_L^{\top}\tilde{\boldsymbol{K}}_{\text{L}}^{\text{cmp}}
\right), 
\quad \boldsymbol{S}_L^{cmp}\in\mathbb{R}^{M}
\end{equation}
Let $l'$ denote the selection block size. When the compression and selection block schemes differ, LTIS derives the score of each selection block from the compressed-block scores according to their spatial overlap. Given $l \le l'$, $s \mid l$, and $s \mid l'$, the selection score for block $j$ is computed as:
\begin{equation}
\boldsymbol{S}^{LTIS}_{\text{L}}[j]=
\sum_{m=0}^{\frac{l'}{s}-1}\;
\sum_{n=0}^{\frac{l}{s}-1}
\boldsymbol{S}^{cmp}_{\text{L}}\!\left[\frac{l'}{s}\,j-m-n\right]
\label{eq:slc-from-cmp}
\end{equation}
here, $[\,\cdot\,]$ denotes the indexing operator for accessing a vector element. For models employing GQA, where key--value caches are shared across query heads, the importance scores are also shared across heads within the same group. The group-shared importance scores are defined as:
\begin{equation}
    \boldsymbol{S}_{\mathrm{L}}^{\mathrm{LTIS},g}
    = \sum_{h=1}^{H_g} \boldsymbol{S}_{\mathrm{L}}^{\mathrm{LTIS},(g,h)}
    \label{eq:group-slc}
\end{equation}
where $g$ is the KV group, $h$ indexes query heads within it, and $H_g$ is the number of query heads per group.

\noindent \textbf{Top-$k$ Selection.} For each query position and each attention head or GQA group, LTIS selects the top-$k$ key--value blocks with the largest selection scores:
\begin{equation}
    \mathcal{I}_{t,g}
    = \operatorname{TopK}_{k}\!\left(\boldsymbol{S}_{\mathrm{L}}^{\mathrm{LTIS},g}[t,:]\right)
    \label{eq:topk-selection}
\end{equation}
where $\mathcal{I}_{t,g}$ is the selected block-index set for query position $t$ and group $g$. The selected keys are concatenated as
\begin{equation}
    \tilde{\boldsymbol{K}}_{\mathrm{L},t,g}^{\mathrm{LTIS}}
    = \operatorname{Cat}\!\left\{
    \boldsymbol{K}_{i l'+1:(i+1)l'} \mid i \in \mathcal{I}_{t,g}
    \right\}
    \in \mathbb{R}^{kl' \times d_k}
    \label{eq:selected-keys}
\end{equation}

The same operation is applied to obtain $\tilde{\boldsymbol{V}}_{\mathrm{L},t,g}^{\mathrm{LTIS}}$. This block selection procedure can be implemented with Triton~\cite{tillet2019triton} kernels, following efficient block-sparse attention implementations such as Native Sparse Attention (NSA)~\cite{yuan2025native}.

\subsection{Short-Term Interest Selection (STIS)}
\label{sec:STIS}
Short-term interests are transient and often evolve within a narrow temporal horizon~\cite{shen2022hierarchically}. STIS is designed to capture such local preference from recent interactions. Prior studies have shown that recent interactions are important for sequential recommendations~\cite{liu2023linrec,shen2022hierarchically,li2023strec}. Therefore, STIS encourages each interaction to attend primarily to local neighborhoods, enabling the model to learn local transition patterns with reduced attention computation.
 
To this end, STIS adopts a power-law mask~\cite{chen2025powerattention,li2019logsparse} in the attention computation. Specifically, each interaction is allowed to attend to: (i) its direct neighbors and (ii) interactions located at distances corresponding to integer powers of two. This pattern provides a more compact receptive field than sliding-window attention (SWA)~\cite{SWA}, while reducing computational complexity~\cite{chen2025powerattention,li2019logsparse}. When sequences are exceptionally long, blocks can be treated as the basic units for applying sparse attention patterns~\cite{beltagy2020longformer,zaheer2020big,guo2019star,jiang2024minference,chen2025powerattention}.

Furthermore, to amplify the influence of the most recent interactions on the next action, each query block is required to attend to every position in the most recent block. This inductive bias explicitly injects recent behavioral signals into the learned representations, strengthening the connection between imminent and historical interactions without sacrificing the overall sparsity of the attention pattern.

\noindent \textbf{Power Attention Mask.}
For short-term interest selection, we introduce a mask that allows a query position to attend to:
(1) interactions within a symmetric local window of 
$w = \texttt{win}\times\texttt{blk}$ interactions;
(2) whole blocks whose block-index distance from the query block is a power of two, 
i.e., $|\,b_q - b_k\,| = 2^r,\; r\in\mathbb N$, and
(3) the final block, which preserves the most recent interactions. Here, $\texttt{win}$ denotes the window size in terms of interaction blocks, and $\texttt{blk}$ denotes the block size. Let $b_q = \bigl\lfloor i/\texttt{blk}\bigr\rfloor$ and $b_k = \bigl\lfloor j/\texttt{blk}\bigr\rfloor$ be the block indices of query position $i$ and key position $j$, respectively. The attention mask $\boldsymbol M_{\text{STIS}}\in\{0,1\}^{L\times L}$ is defined as:
\begin{equation}
\boldsymbol M_{\text{STIS}}(i,j)=
\begin{cases}
1, & |\,i-j\,|< w \\[2pt]
1, & |\,b_q - b_k\,|=2^k \text{ for some } k\in\mathbb N \\[2pt]
1, & j \text{ belongs to the last } \texttt{blk} \text{ positions} \\[2pt]
0, & \text{otherwise}
\end{cases}
\end{equation}

This design achieves logarithmic sparse connectivity while maintaining a receptive field that grows logarithmically with the sequence length. The forced visibility of the last block aligns with empirical findings on the importance of recent interactions in sequential recommendation.

\subsection{Learnable Output Gating}
\label{sec:MLP}
Effectively fusing the two attention outputs is critical for balancing long-term and short-term user interests. A naive additive combination may cause the fused output to deviate substantially from the standard attention output. Since the two pathways capture complementary aspects of user interests, a weighted aggregation is preferable. Therefore, we introduce a learnable gating mechanism whose weights are produced by a learnable MLP followed by a sigmoid activation:
\begin{equation}
    \boldsymbol{O}_{\text{LTIS}} =
    \operatorname{GQA}
    \left(
    \boldsymbol{Q},
    \tilde{\boldsymbol{K}}_{\text{L}}^{\text{LTIS}},
    \tilde{\boldsymbol{V}}_{\text{L}}^{\text{LTIS}}
    \right)
\end{equation}
\begin{equation}
    \boldsymbol{O}_{\text{STIS}} =
    \operatorname{GQA}
    \left(
    \boldsymbol{Q},
    \boldsymbol{K},
    \boldsymbol{V},
    \boldsymbol{M}_{\text{STIS}}
    \right)
\end{equation}
\begin{equation}
    \boldsymbol{\alpha}
    =
    \sigma
    \left(
    \mathcal{F}_{o}
    \left(
    [\boldsymbol{O}_{\text{LTIS}};\boldsymbol{O}_{\text{STIS}}]
    \right)
    \right)
\end{equation}
where $\mathcal{F}_{o}$ is a learnable MLP, $\boldsymbol{O}_{\text{LTIS}}$ and $\boldsymbol{O}_{\text{STIS}}$ are the attention outputs of LTIS and STIS, respectively, $\sigma$ denotes the sigmoid activation function, and $\boldsymbol{\alpha} \in [0,1]$ represents the gating weights. The final attention output is computed as:
\begin{equation}
    \boldsymbol{O}_{\text{Blossom}}
    =
    \boldsymbol{\alpha} \odot \boldsymbol{O}_{\text{LTIS}}
    +
    (1 - \boldsymbol{\alpha}) \odot \boldsymbol{O}_{\text{STIS}}
\end{equation}
where $\odot$ denotes element-wise multiplication. This adaptive fusion allows the model to learn context-dependent attention strategies.

The output of Blossom Attention is then fed into the Transformer layer:
\begin{equation}
    \boldsymbol{S}^{n}
    =
    \operatorname{LayerNorm}
    \left(
    \boldsymbol{H}^{n-1}
    +
    \operatorname{Dropout}
    \left(
    \boldsymbol{O}_{\text{Blossom}}
    \left(
    \boldsymbol{H}^{n-1}
    \right)
    \right)
    \right)
\end{equation}
\begin{equation}
    \boldsymbol{H}^{n}
    =
    \operatorname{LayerNorm}
    \left(
    \boldsymbol{S}^{n}
    +
    \operatorname{Dropout}
    \left(
    \operatorname{FFN}
    \left(
    \boldsymbol{S}^{n}
    \right)
    \right)
    \right)
\end{equation}
\begin{equation}
    \boldsymbol{H}^{0} = \boldsymbol{E}, 
    \quad
    \boldsymbol{H} = \boldsymbol{H}^{N}\boldsymbol{W}_{N} + \boldsymbol{b}_{N}
\end{equation}
where $\operatorname{LayerNorm}(\cdot)$ denotes layer normalization~\cite{ba2016layer}, and $\boldsymbol{H}^{n}$ is the hidden state at the $n$-th layer, where $n = 1, \cdots, N$. The function $\operatorname{FFN}(\cdot)$ is the feed-forward network. The parameters $\boldsymbol{W}_{N} \in \mathbb{R}^{d \times d}$ and $\boldsymbol{b}_{N} \in \mathbb{R}^{d}$ denote the output projection weight and bias, respectively. Finally, we obtain the sequence representation $\boldsymbol{H} \in \mathbb{R}^{L \times d}$. The inference and optimization details are provided in Appendix~\ref{sec:IO}.

\subsection{In-Depth Analysis}
\label{sec:complexity}
We analyze the theoretical computational complexity of BlossomRec for a sequence of length $L$ and embedding dimension $d$.

\noindent \textbf{LTIS Complexity.}
The block partitioning step creates $M=\Bigl\lfloor\dfrac{L-l}{s}\Bigr\rfloor+1$ blocks. The importance scoring step has a complexity of $O(M^2d)$, as attention is computed over the $M$ compressed blocks. The attention computation has a complexity of $O((l'k)^2d)$. With GQA, the corresponding complexity becomes $O(G(l'k)^2d)$, where $G$ is the number of key--value groups. Thus, the total LTIS complexity is:
\begin{equation}
    \mathcal{O}_{\text{LTIS}} = O\left(M^2d + G(l'k)^2d\right)
\end{equation}

Since $s \ll L$ and $l'k$ is relatively small, the complexity is approximately $O((L/s)^2d)$, which is significantly lower than the $O(L^2d)$ complexity of standard full attention.
  
\noindent \textbf{STIS Complexity.}
The power-law mask introduces $O(\log(L/b))$ visible blocks, where $b$ is the block size. Therefore, the total STIS complexity is:
\begin{equation}
    \mathcal{O}_{\text{STIS}} = O\left(\log(L/b)d\right)
\end{equation}
    
\noindent \textbf{Overall Complexity.}
Since the gating mechanism introduces only negligible additional parameters, with complexity $O(d)$, the overall theoretical complexity is:
\begin{equation}
    \mathcal{O}_{\text{Blossom}} =
    O\left(M^2d + G(l'k)^2d + \log(L/b)d\right)
\end{equation}
    
With appropriate hyperparameter settings, where $s \ll L$, $b \ll L$, and $l'k \ll L$, BlossomRec achieves sub-quadratic complexity while maintaining expressive modeling capacity. A case study of the efficiency analysis is provided in Appendix~\ref{sec:E Ana}.

\begin{table}[t]
\centering
\caption{Statistics of the datasets.}
\label{dataset}
\begin{tabular}{lrrrr}
\toprule
\textbf{Dataset} & \#Users & \#Items & \#Inters & Sparsity \\
\midrule
ML-1M & 6,041 & 3,707 & 1,000,209 & 95.53\% \\
Gowalla & 64,116 & 164,533 & 2,018,421 & 99.98\% \\
Amazon Video Games & 94,763 & 25,613 & 814,586 & 99.97\% \\
Amazon Beauty & 22,364 & 12,102 & 198,502 & 99.93\% \\
\bottomrule
\end{tabular}
\end{table}

%% file: 4Experiment.tex
\begin{table*}[t]
    \caption{Overall performances on four datasets. All improvements are
statistically significant (i.e., two-sided t-test with p < 0.05) over baseline models, except for Recall@10 on ML-1M compared with SASRec and MRR@10 on Gowalla compared with LinRec. In each row, the best result is bold, while the second-best result is underlined.}
    \label{tab:overall-performance}
    \begin{tabular}{l|c|ccccc|c}
        \toprule
        \textbf{Dataset} & \textbf{Metric} & SASRec & BERT4Rec & GRU4Rec & LinRec & Mamba4Rec & \textbf{BlossomRec} \\
        \midrule
        \multirow{3}{*}{ML-1M} 
         & Recall@10 & \textbf{0.8152} & 0.8088 & 0.7987 & 0.8113 & 0.8116 & \uline{0.8151} \\
         & MRR@10 & 0.5442 & 0.5338 & 0.5274 & 0.5431 & \uline{0.5472} & \textbf{0.5485} \\
         & NDCG@10 & 0.6097 & 0.6004 & 0.5928 & 0.6078 & \uline{0.6111} & \textbf{0.6128} \\
        \midrule
        \multirow{3}{*}{Gowalla} 
         & Recall@10 & 0.9428 & 0.9278 & 0.9392 & \uline{0.9441} & 0.9424 & \textbf{0.9482} \\
         & MRR@10 & 0.7739 & 0.7409 & 0.7587 & \textbf{0.7793} & 0.7763 & \uline{0.7781} \\
         & NDCG@10 & 0.8154 & 0.7867 & 0.8029 & \uline{0.8198} & 0.8172 & \textbf{0.8200} \\
        \midrule
        \multirow{3}{*}{Amazon Video Games} 
         & Recall@10 & 0.7372 & 0.6973 & 0.7257 & \uline{0.7375} & 0.7256 & \textbf{0.7380} \\
         & MRR@10 & 0.4656 & 0.4218 & 0.4513 & \uline{0.4667} & 0.4537 & \textbf{0.4671} \\
         & NDCG@10 & 0.5304 & 0.4874 & 0.5168 & \uline{0.5313} & 0.5186 & \textbf{0.5317} \\
        \midrule
        \multirow{3}{*}{Amazon Beauty} 
         & Recall@10 & \uline{0.4723} & 0.4039 & 0.4679 & 0.4674 & 0.4298 & \textbf{0.4734} \\
         & MRR@10 & 0.2774 & 0.2128 & 0.2644 & \uline{0.2810} & 0.2427 & \textbf{0.2842} \\
         & NDCG@10 & 0.3235 & 0.2578 & 0.3124 & \uline{0.3251} & 0.2868 & \textbf{0.3289} \\
        \bottomrule
    \end{tabular}
\end{table*}

\section{Experiments}

In this section, we present extensive experimental results to validate the effectiveness of BlossomRec on SR tasks. The following \textbf{R}esearch \textbf{Q}uestions will be answered by analysis of the experimental results:
\begin{itemize}[leftmargin=*]
    \item \textbf{RQ1}: How does BlossomRec perform when integrating with other transformer-based models and compared with other state-of-the-art SRS models?
    \item \textbf{RQ2}: How does BlossomRec perform in terms of efficiency?
    \item \textbf{RQ3}: What is the influence on the performance of the core components in BlossomRec?
    \item \textbf{RQ4}: How do the hyperparameters influence the performance of BlossomRec?
    \item \textbf{RQ5}: Why can BlossomRec improve recommendation performance?
\end{itemize}

\subsection{Experiment Settings}
\subsubsection{\textbf{Datasets}} We conduct experiments on four widely-used benchmark datasets: MovieLens-1M (ML-1M)\footnote{https://grouplens.org/datasets/movielens/}, Gowalla\footnote{https://snap.stanford.edu/data/loc-gowalla.html}, Amazon Games, and Amazon Beauty\footnote{http://jmcauley.ucsd.edu/data/amazon/}. 
\textbf{ML-1M} contains 1 million movie ratings.
\textbf{Gowalla} is a location-based social network dataset with check-in records. 
\textbf{Amazon Video Games} and \textbf{Amazon Beauty} are subsets of the Amazon product dataset,
These datasets represent diverse application scenarios with varying characteristics. Following other previous work \cite{liu2025sigma, liu2023linrec, kang2018self}, we set the settings for the datasets. The datasets' statistics are as in Table \ref{dataset}.

\subsubsection{\textbf{Evaluation Metrics}} 
We adopt the leave-one-out evaluation strategy. Specifically, for each user sequence, we hold out the last interaction for testing, use the second-to-last interaction for validation, and use the remaining interactions for training. We evaluate model performance using three ranking metrics: Recall@10, MRR@10 (Mean Reciprocal Rank), and NDCG@10 (Normalized Discounted Cumulative Gain).

\subsubsection{\textbf{Baselines}} We compare BlossomRec with several state-of-the-art SRS models: 
(1) \textbf{GRU4Rec}~\cite{jannach2017recurrent}, a RNN-based method for session-based recommendation; (2) \textbf{SASRec}~\cite{kang2018self}, a self-attention based SR model; 
(3) \textbf{BERT4Rec}~\cite{sun2019bert4rec}, which employs bidirectional self-attention with Cloze task for SR; 
(4) \textbf{LinRec}~\cite{liu2023linrec}, an efficient linear-complexity SR model; 
(5) \textbf{Mamba4Rec}~\cite{liu2024mamba4rec}, a SSM-based approach for SR.

\subsubsection{\textbf{Implementation Details}} \label{sec:expset}
All models are implemented using PyTorch 2.6, Triton 3.2, Python 3.12, and RecBole 1.2.1\footnote{\url{https://recbole.io/}}, and are trained on an NVIDIA RTX 4090 GPU. Additional implementation details are provided in Appendix~\ref{sec:I Detail}.

\begin{table}[t]
    \centering
    \caption{Results of transferability experiment. ``w/o'' denotes the original backbone, ``w'' denotes equipping it with our BlossomRec attention.}
    \label{tab:blossom-ablation}
    \begin{tabular}{lcccccc}
        \toprule
        \multirow{2}{*}{Dataset} & \multirow{2}{*}{Metric} & \multicolumn{2}{c}{SASRec} & \multicolumn{2}{c}{BERT4Rec} \\
        \cmidrule(lr){3-4} \cmidrule(lr){5-6}
         & & w/o & w & w/o & w \\
        \midrule
        \multirow{3}{*}{ML-1M} 
         & Recall@10 & \textbf{0.8152} & {0.8151} & 0.8088 & 0.8033 \\
         & MRR@10 & 0.5442 & \textbf{0.5485} & 0.5338 & 0.5409 \\
         & NDCG@10 & 0.6097 & \textbf{0.6128} & 0.6004 & 0.6043 \\
        \midrule
        \multirow{3}{*}{Gowalla} 
         & Recall@10 & 0.9428 & \textbf{0.9482} & 0.9278 & 0.9304 \\
         & MRR@10 & 0.7739 & \textbf{0.7781} & 0.7409 & 0.7487 \\
         & NDCG@10 & 0.8154 & \textbf{0.8200} & 0.7867 & 0.7933 \\
        \midrule
        \multirow{3}{*}{Games} 
         & Recall@10 & 0.7372 & \textbf{0.7380} & 0.6973 & 0.6959 \\
         & MRR@10 & 0.4656 & \textbf{0.4671} & 0.4218 & 0.4208 \\
         & NDCG@10 & 0.5304 & \textbf{0.5317} & 0.4874 & 0.4864 \\
        \midrule
        \multirow{3}{*}{Beauty} 
         & Recall@10 & {0.4723} & \textbf{0.4734} & 0.4039 & 0.4179 \\
         & MRR@10 & 0.2774 & \textbf{0.2842} & 0.2128 & 0.2200 \\
         & NDCG@10 & 0.3235 & \textbf{0.3289} & 0.2578 & 0.2665 \\
        \bottomrule
    \end{tabular}
\end{table}

\subsection{Overall Performance (RQ1)}

Table \ref{tab:overall-performance} presents the comprehensive comparison between BlossomRec and baseline models across four datasets.

\begin{itemize}[leftmargin=*]
    \item The experimental results demonstrate that BlossomRec achieves superior or competitive performance across all datasets. BlossomRec achieves the best performance on 10 out of 12 metrics (marked in bold). Notably, on the Gowalla dataset, BlossomRec improves Recall@10 by 0.54 percentage points (from 0.9428 to 0.9482) and NDCG@10 by 0.46 percentage points (from 0.8154 to 0.8200) compared to SASRec. Similar improvements are observed on Amazon Beauty, where MRR@10 increases by 0.68 percentage points (from 0.2774 to 0.2842).
    \item Compared to GRU4Rec, LinRec and Mamba4Rec, BlossomRec shows stronger performance on most datasets, indicating that our carefully designed architecture better captures user interests.
    \item The result from Table~\ref{tab:blossom-ablation} is that BlossomRec achieves comparable or superior performance to the transformer based models using standard attention, demonstrating its strong transferability. This validates the superiority of BlossomRec.
\end{itemize}

\subsection{Efficiency Study (RQ2)}
\begin{figure}[t]
\vspace{-2mm}
\centering
{\includegraphics[width=.48\linewidth]{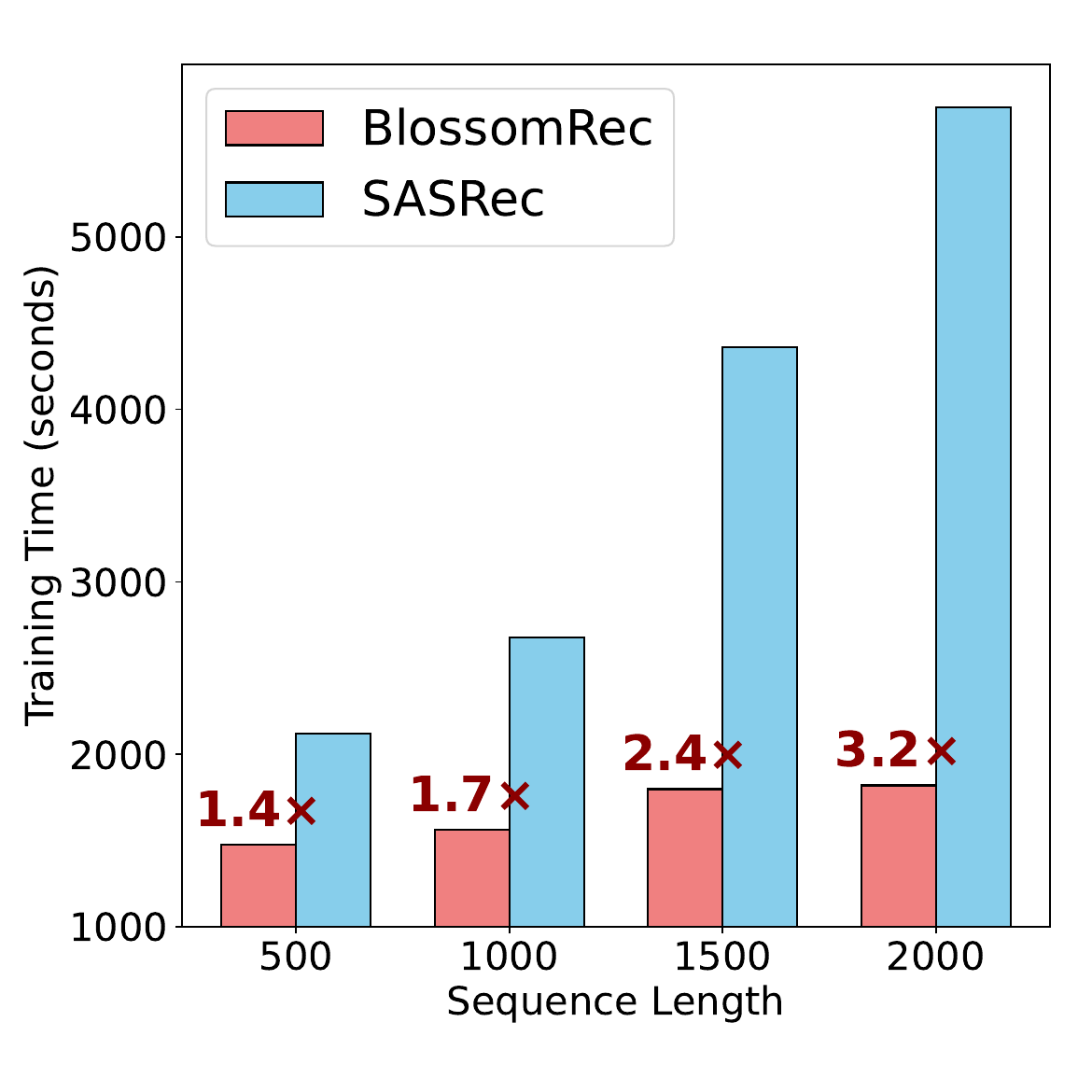}}\hfill
{\includegraphics[width=.48\linewidth]{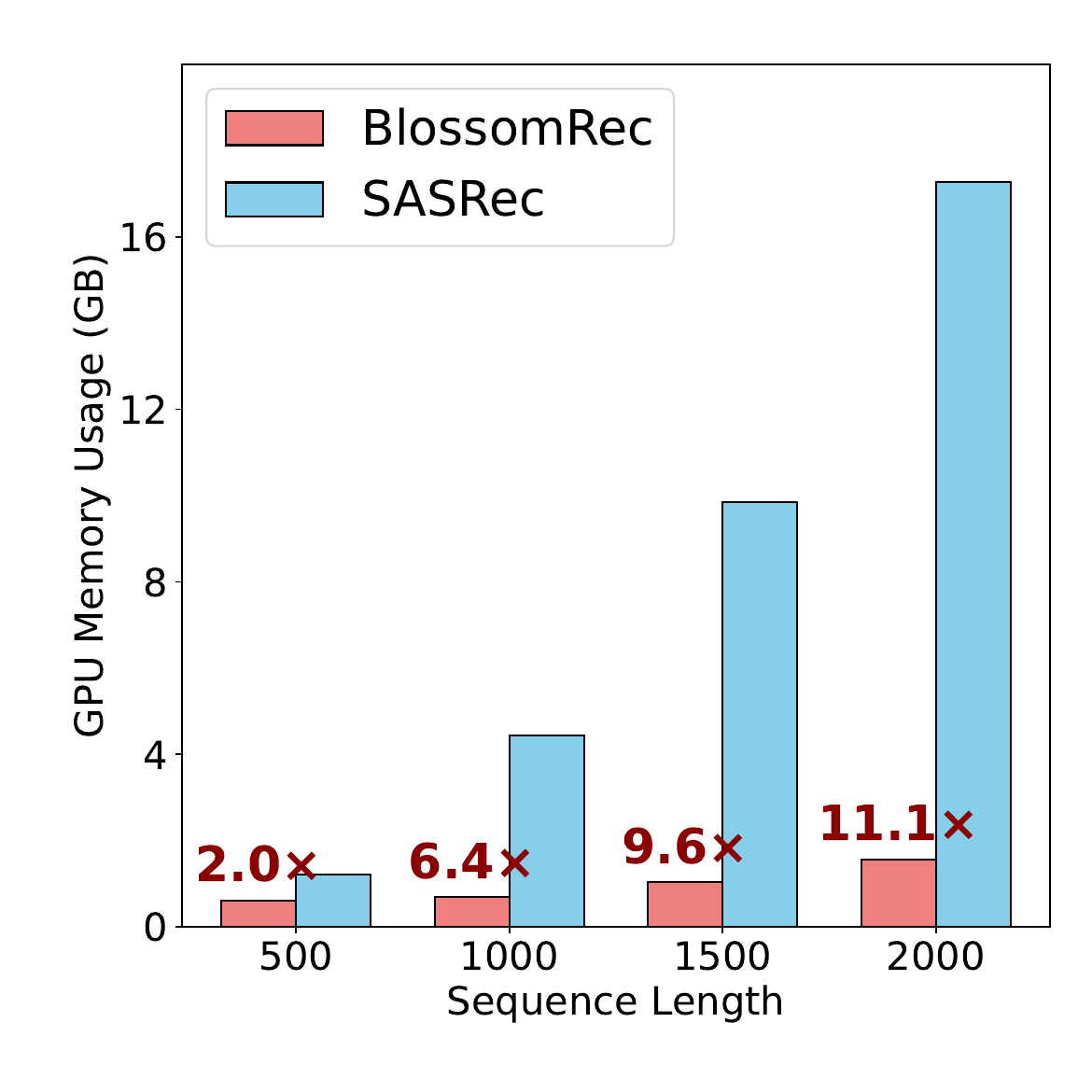}}\\[-2mm]   
{\includegraphics[width=.48\linewidth]{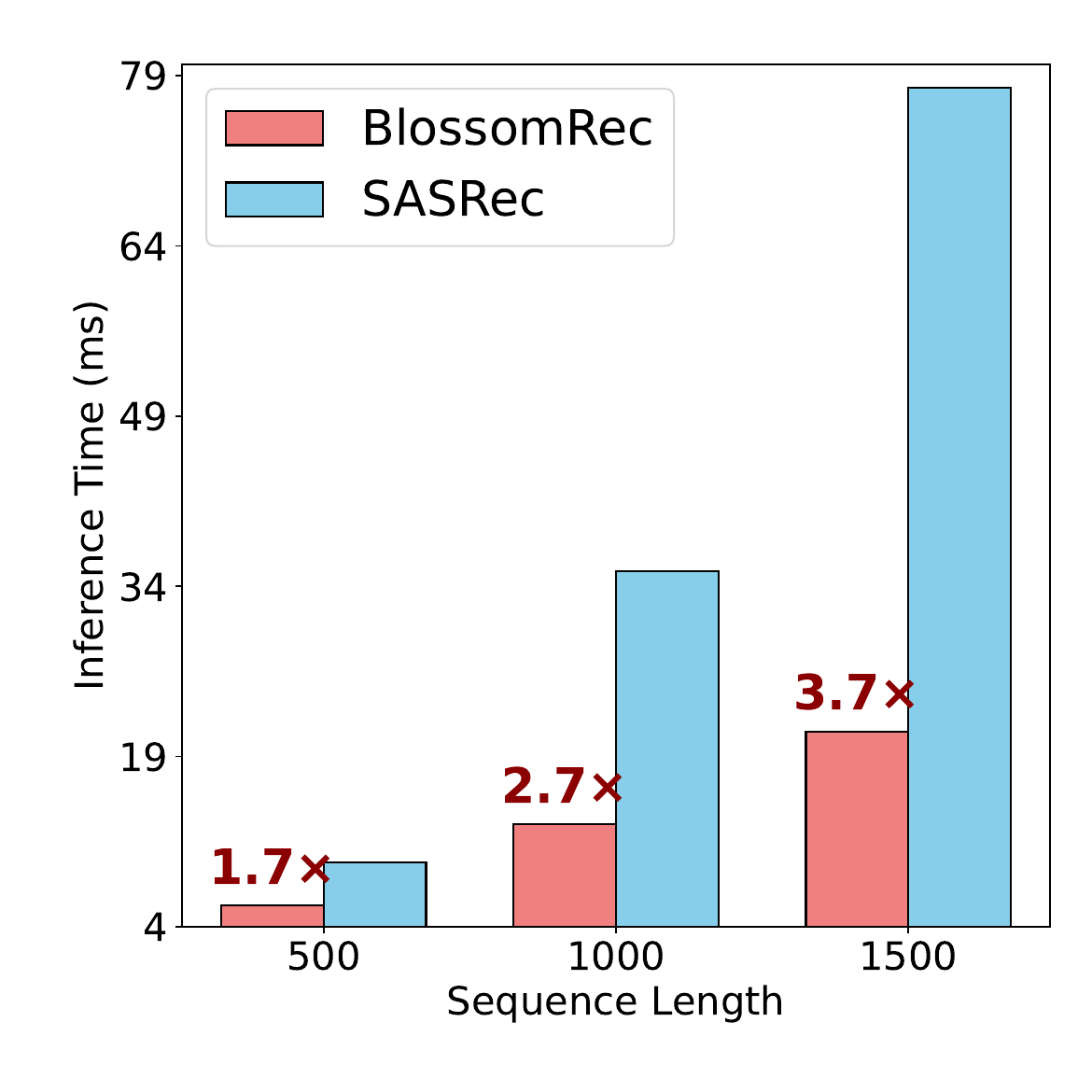}}\hfill
{\includegraphics[width=.48\linewidth]{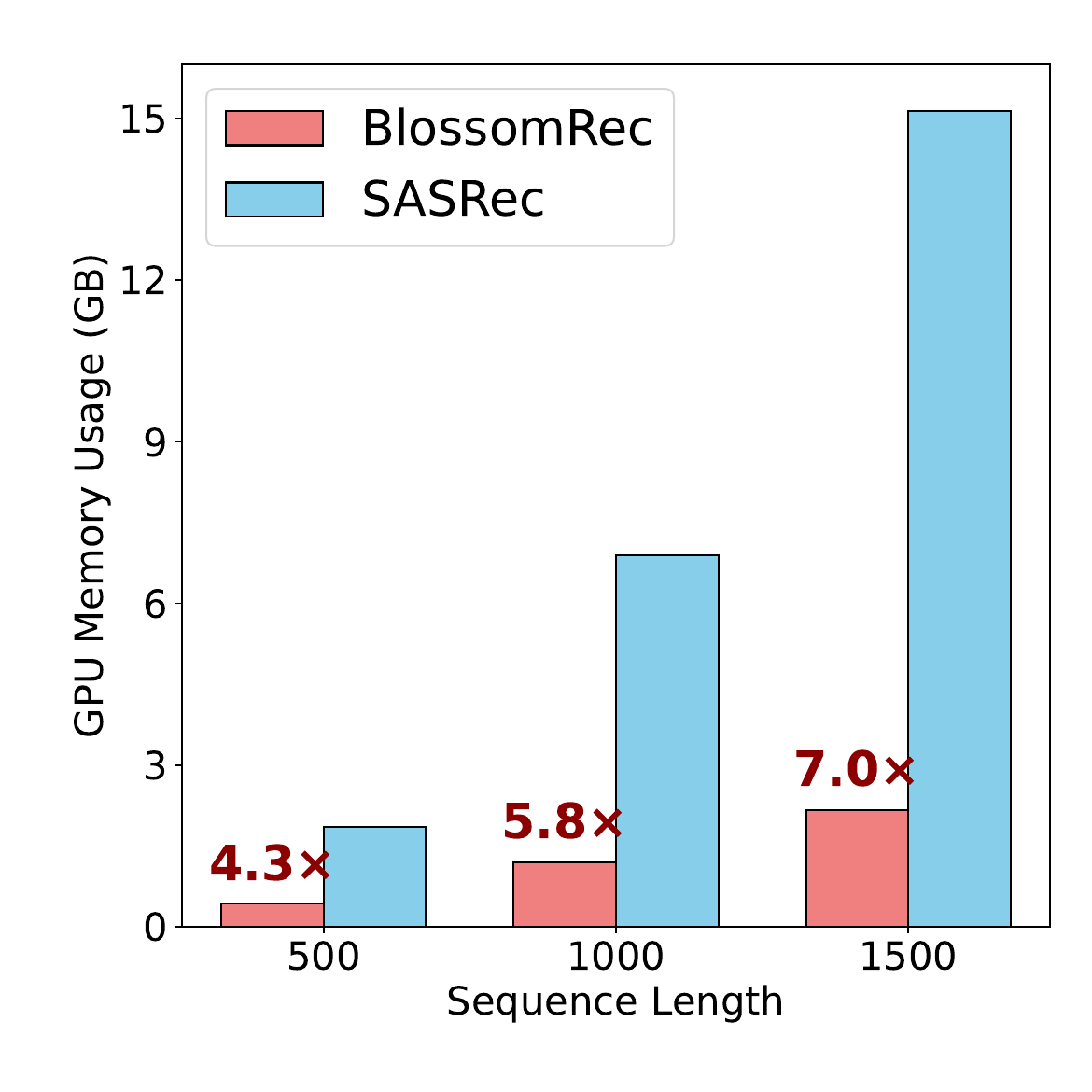}}
\vspace{-2mm}
\caption{Efficiency study results.}
\label{fig:efficiency}
\end{figure}

\noindent Figure~\ref{fig:efficiency} shows the training and inference efficiency results of BlossomRec compared to SASRec under different sequence lengths. Detailed experimental configurations are provided in Appendix~\ref{sec:efficienset}.

\begin{itemize}[leftmargin=*]
    \item \textbf{Training Efficiency.} Training time in Figure~\ref{fig:efficiency}~(upper-left) exhibits a fast escalation for SASRec, whereas BlossomRec scales slowly: at a sequence length of 2,000, BlossomRec completes an epoch 3.2× faster. This acceleration stems from the reduced computational complexity of block-wise sparse attention.
    GPU memory footprint during training in Figure~\ref{fig:efficiency}~(upper-right) widens in favor of BlossomRec as sequences lengthen. At length 2,000, BlossomRec consumes approximately one eleventh of the GPU memory required by SASRec. The sparse attention mechanism alleviates the heavy computation, making training on long sequences friendly.
    \item \textbf{Inference Efficiency.} Inference time in Figure~\ref{fig:efficiency}~(lower-left) scales slowly for BlossomRec, while SASRec exhibits a steep rise. Consequently, at length 1,500, BlossomRec attains a 3.7× speed-up over SASRec, a critical advantage for latency-sensitive recommendation services.
    GPU memory usage during inference in Figure~\ref{fig:efficiency}~(lower-right) follows a similar pattern: BlossomRec consistently demands less memory, and the gap magnifies with sequence length. At length 1,500, its GPU memory requirement is roughly one-seventh of SASRec’s. The sparsity structure, therefore, alleviates not only computational but also memory bottlenecks at serving time, facilitating deployment in resource-constrained environments.
\end{itemize}

\subsection{Ablation Study (RQ3)}
\begin{table}[t]
\centering
\vspace{-2mm}
\caption{Ablation study results on ML-1M.}
\vspace{-2mm}
\label{tab:ablation_ml1m}
\begin{tabular}{lccc}
\toprule
\textbf{Method} & \textbf{Recall@10} & \textbf{MRR@10} & \textbf{NDCG@10} \\
\midrule

LTIS-only & 0.8094 & 0.5336 & 0.6001 \\
STIS-only & 0.8093 & 0.5441 & 0.6081 \\
LTIS+SWA & \uline{0.8109} & \uline{0.5454} & \uline{0.6095} \\
\hline
\textbf{BlossomRec} & \textbf{0.8151} & \textbf{0.5485} & \textbf{0.6128} \\
\bottomrule
\end{tabular}
\vspace{-3mm}
\end{table}
 
\noindent Table~\ref{tab:ablation_ml1m} summarizes the contribution of each core submodule in the proposed BlossomRec architecture. Four controlled variants are evaluated: (1) \emph{LTIS-only}, where the STIS branch is disabled. (2) \emph{STIS-only}, where the LTIS branch is disabled. (3) \emph{LTIS + SWA}, where the power-law mask in STIS is replaced with sliding-window attention (SWA) using a fixed window size of 16. (4) \emph{BlossomRec}, the full model.

\begin{itemize}[leftmargin=*]
    \item Disabling either branch consistently hurts performance, confirming the necessity of the parallel sparse-attention design. The \emph{LTIS-only} variant suffers the largest decline, with relative drops of 0.70\% in Recall@10, 2.72\% in MRR@10, and 2.07\% in NDCG@10. This suggests that short-term interests play an important role in recommendation performance. Conversely, the \emph{STIS-only} variant degrades NDCG@10 by 0.77\%, indicating that long-term interest modeling also contributes to the final performance.
    \item Replacing the STIS branch with sliding-window attention (\emph{LTIS + SWA}) restricts the receptive field to a fixed local window. Although this variant outperforms the single-branch ablations, it still trails the full model by 0.52\% in Recall@10 and 0.54\% in NDCG@10, confirming that the power-law attention mask is more suitable for SRS.
    \item In summary, these results verify that the dual-branch design is not merely an additive combination of two sparse attention patterns. Instead, the LTIS and STIS pathways function complementarily by modeling long-term and short-term user interests, respectively, thereby achieving the best overall performance.
\end{itemize}

\subsection{Parameter Analysis (RQ4)}
\begin{figure}[t]
\vspace{-2mm}
\centering

{\includegraphics[width=.48\linewidth]{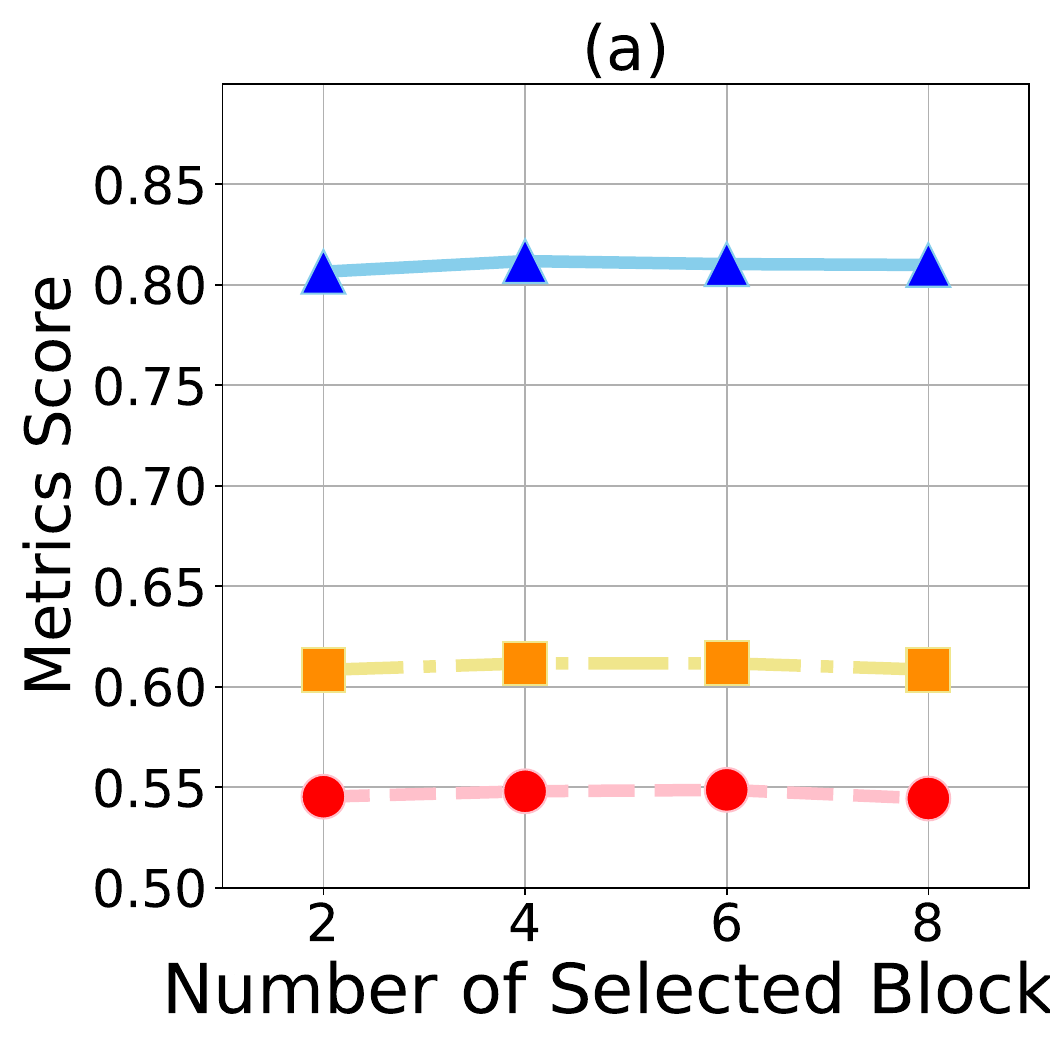}}\hfill
{\includegraphics[width=.48\linewidth]{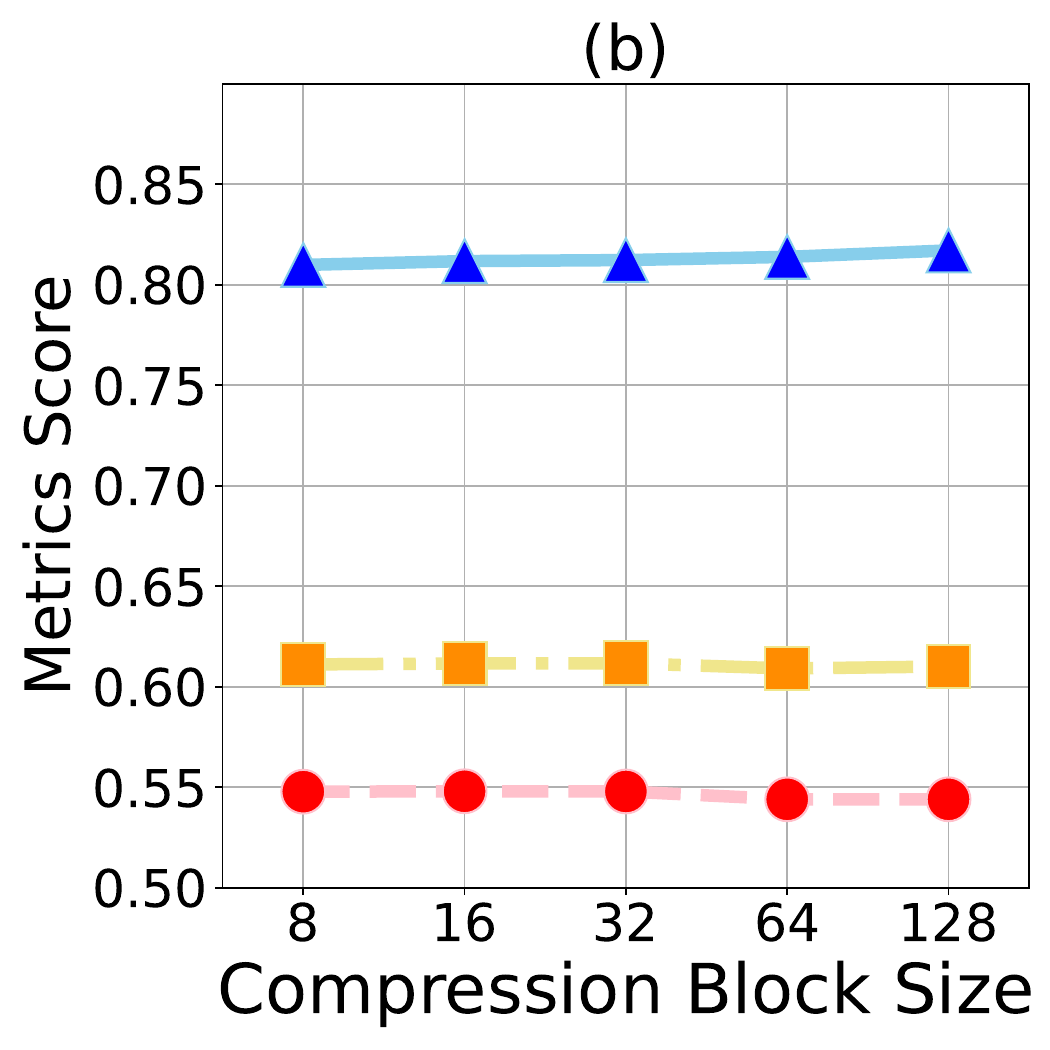}}\\[0.mm]   
{\includegraphics[width=.48\linewidth]{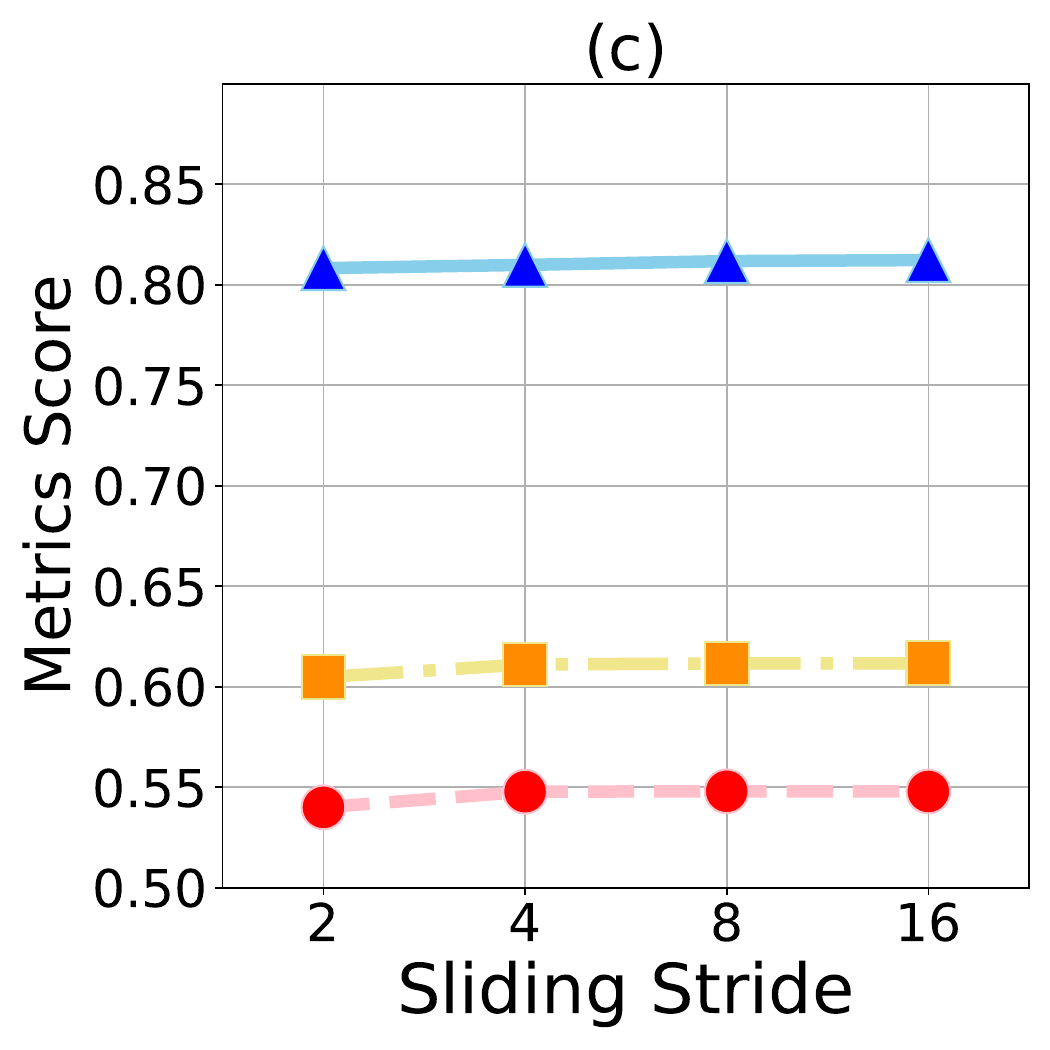}}\hfill
{\includegraphics[width=.48\linewidth]{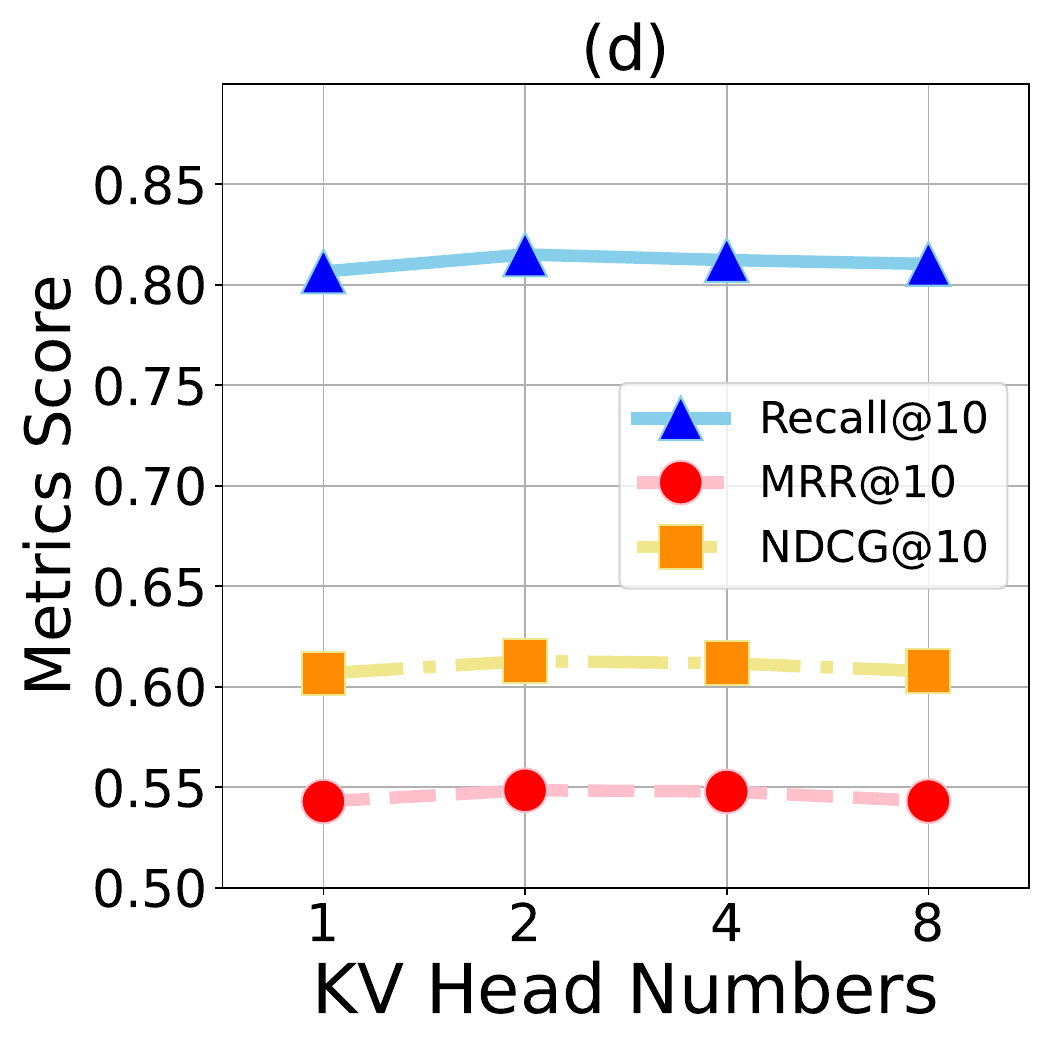}}
\vspace{-2mm}
\caption{Parameter analysis results.}
\vspace{-6mm}
\label{fig:Parameter}
\end{figure}

The selection block size $l'$ in Equation~\ref{eq:slc-from-cmp} is set to 16 to satisfy the minimum requirement of the Triton operator. Figure~\ref{fig:Parameter}(a) illustrates the variation of the three metrics with respect to the number of selected blocks. Recall@10 reaches its peak when the number of selected blocks is 4, while both MRR@10 and NDCG@10 achieve better performance at values of 4 and 6 than under other settings. 
Figure~\ref{fig:Parameter} (b) shows that Recall@10 increases monotonically with the compression size. MRR@10 exhibits comparable performance for compression sizes of 8, 16, and 32, all of which outperform the results obtained with compression sizes of 64 and 128. NDCG@10 reaches its best performance at compression sizes of 16 and 32, with 32 yielding well-balanced overall performance across all metrics. 
Figure~\ref{fig:Parameter} (c) shows that all three metrics gradually improve as the sliding stride increases, with Recall@10 reaching its maximum at a stride of 16. Figure~\ref{fig:Parameter} (d) indicates that all three metrics attain their highest values when the number of KV heads is set to 2, corresponding to 4 GQA groups. 
For the parameters in STIS, \texttt{blk} should not be set excessively large, as doing so may cause nearly the entire sequence to participate in attention computation. The parameter \texttt{win} denotes the window size in the power-law mask. Experimental results show that a window size of 8, with NDCG@10 = 0.6128, outperforms a window size of 4, with NDCG@10 = 0.6101.

\subsection{Case Study (RQ5)}

\begin{figure}[t]
\centering
\vspace{-3mm}
\includegraphics[height=3.23cm]{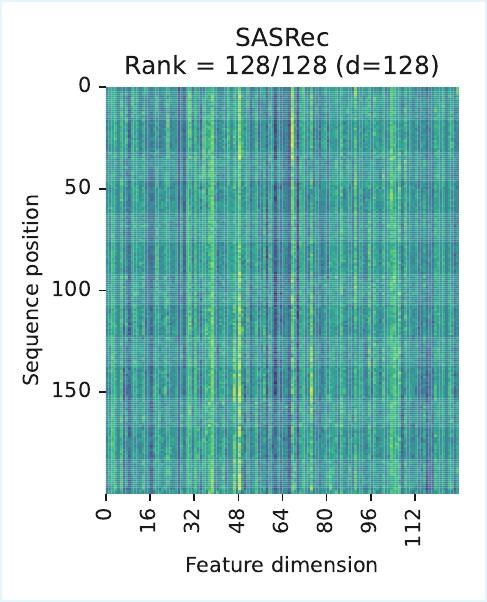}\hfill
\includegraphics[height=3.23cm]{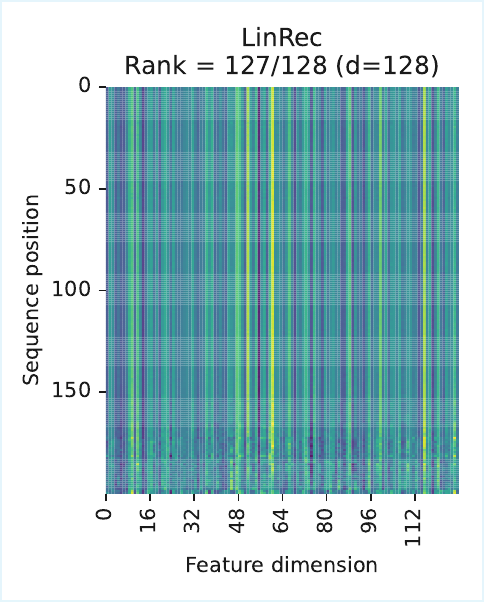}\hfill
\includegraphics[height=3.23cm]{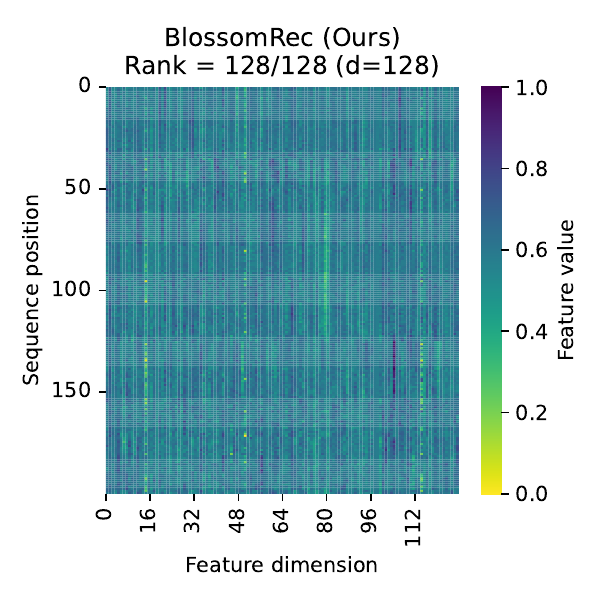}
\vspace{-4mm}
\caption{Feature-map visualization of different models.}
\vspace{-4mm}
\label{fig:feature_map}
\end{figure}

As illustrated in Figure~\ref{fig:feature_map}, we manually select one interaction sequence from ML-1M and visualize the final-layer feature maps produced by three SR models. BlossomRec preserves full-rank feature representations, indicating its ability to retain richer feature information. Moreover, the feature values produced by BlossomRec are generally larger than those of SASRec and LinRec, suggesting stronger global representation capacity. In contrast, the feature map generated by LinRec is lighter and concentrates primarily on the last few positions, while earlier positions exhibit highly similar patterns, revealing a strong recency bias. Although SASRec and BlossomRec produce feature maps with similar overall structures and both capture features across the entire sequence, BlossomRec exhibits richer and more diverse textural patterns, suggesting more stable and expressive representations. We attribute this improvement to its explicit long- and short-term interest selection mechanism.

%% file: 5Relatedwork.tex
\section{Related Work}
\subsection{Sequential Recommendations} Compared with other recommendation paradigms~\cite{li2025survey,zhao2018deep,zhao2018recommendations,wang2023plate,li2022gromov,fu2023unified,zhang2024notellm,liu2023exploration, li2026collectivekv, zhang2026length}, sequential recommendation (SR) models are primarily constructed using recurrent neural network (RNN) and Transformer architectures. Early RNN-based approaches, such as GRU4Rec~\cite{jannach2017recurrent}, pioneered session-based recommender systems. Subsequently, Transformer-based models have become the dominant paradigm. SASRec~\cite{kang2018self} introduced a self-attention mechanism for sequential recommendation, while BERT4Rec~\cite{sun2019bert4rec} proposed a Cloze task-based approach for bidirectional modeling. However, these methods suffer from the $O({N}^{2})$ computational complexity when processing long sequences, creating a significant computation bottleneck. Our BlossomRec addresses this limitation through a sparse attention mechanism that effectively captures both short-term and long-term user interests while maintaining superior performance and efficiency on long sequences.

Researchers have explored more efficient alternatives to overcome the computational constraints of Transformers. With the advancement of State Space Models (SSMs)~\cite{gu2023mamba} and Recurrent Units, novel mechanisms such as Mamba4Rec~\cite{liu2024mamba4rec}, and RecBLR~\cite{liu2024behavior} have been applied to recommender systems, achieving substantial efficiency improvements. Linear attention-based approaches, such as LinRec~\cite{liu2023linrec}, reduce computational complexity to $\mathcal{O}(N)$ by incorporating linear attention mechanisms into SR. MLP-based models such as SMLP4Rec~\cite{gao2024smlp4rec}, AutoMLP~\cite{li2023automlp} and MLP4Rec~\cite{li2022mlp4rec} have been proposed as efficient alternatives. However, these methods often exhibit unstable performance despite their computational efficiency. In contrast, our BlossomRec maintains consistent performance across sequences of varying lengths while preserving high efficiency on long sequences.

\subsection{Sparse Attention Mechanisms} Recent research has extensively investigated reducing the computational complexity of attention mechanisms. Fixed-pattern sparse attention approaches, such as SWA~\cite{SWA}, LongNet~\cite{ding2023longnet}, LogSparse~\cite{li2019logsparse}, and Power Attention~\cite{chen2025powerattention}, attempt to compute attention scores only at fixed positions during inference. Block-based sparse attention methods improve inference efficiency through Block-based sparse patterns, including MInference~\cite{jiang2024minference} and Block Attention~\cite{ma2024block}. With the popularity of Mixture-of-Experts (MoE) architectures~\cite{liu2024deepseek}, routing-based block attention mechanisms such as MoBA~\cite{lu2025moba}, and NSA~\cite{yuan2025native} have been introduced. Notably, NSA was the first to propose a trainable sparse attention mechanism, inspiring applications like VideoNSA~\cite{song2025videonsa} and MUFASA~\cite{fu2025multimodal}. However, these sparse attention mechanisms often cannot be directly applied to SR due to their high parameter requirements, which may result in marginal efficiency gains and potential accuracy degradation. Our proposed BlossomRec effectively addresses these limitations by maintaining relatively low parameter overhead while applying to various Transformer-based models, ensuring performance and efficiency.

%% file: 6Conclusion.tex
\section{Conclusion}
In this paper, we propose BlossomRec, a novel block-level fused sparse attention mechanism for sequential recommendation that effectively models both long-term and short-term user interests. Specifically, we introduce two parallel sparse attention pathways to efficiently capture distinct user-interest patterns in SRS. Theoretically, BlossomRec can reduce the number of interactions involved in attention computation by nearly 90\% when the sequence length is 2,000. BlossomRec therefore achieves the objective of capturing both long-term and short-term interests in long sequences while maintaining high computational efficiency. Extensive experimental results demonstrate that BlossomRec achieves state-of-the-art performance in SRS, with remarkable computational efficiency on long sequences and a substantially reduced memory footprint. These results validate the effectiveness of the proposed approach in addressing the scalability challenges of SRS, making it suitable for real-world applications involving long user interaction histories.

\begin{acks}
This research was partially supported by National Natural Science Foundation of China (No.62502404), Hong Kong Research Grants Council (Research Impact Fund No.R1015-23, Collaborative Research Fund No.C1043-24GF, General Research Fund No.11218325), Institute of Digital Medicine of City University of Hong Kong (No.9229503), Huawei (Huawei Innovation Research Program), Tencent (Tencent Rhino-Bird Focused Research Program, Tencent University Cooperation Project), Alibaba (CCF-Alimama Tech Kangaroo Fund No. 2024002), Didi (CCF-Didi Gaia Scholars Research Fund), Kuaishou (CCF-Kuaishou Large Model Explorer Fund, Kuaishou University Cooperation Project), and Bytedance.
\end{acks}

%% file: 10Appendix.tex
\section{Observations}
\begin{figure*}[h]
  \centering
  \vspace{-4mm}
  \includegraphics[width=0.8\linewidth]{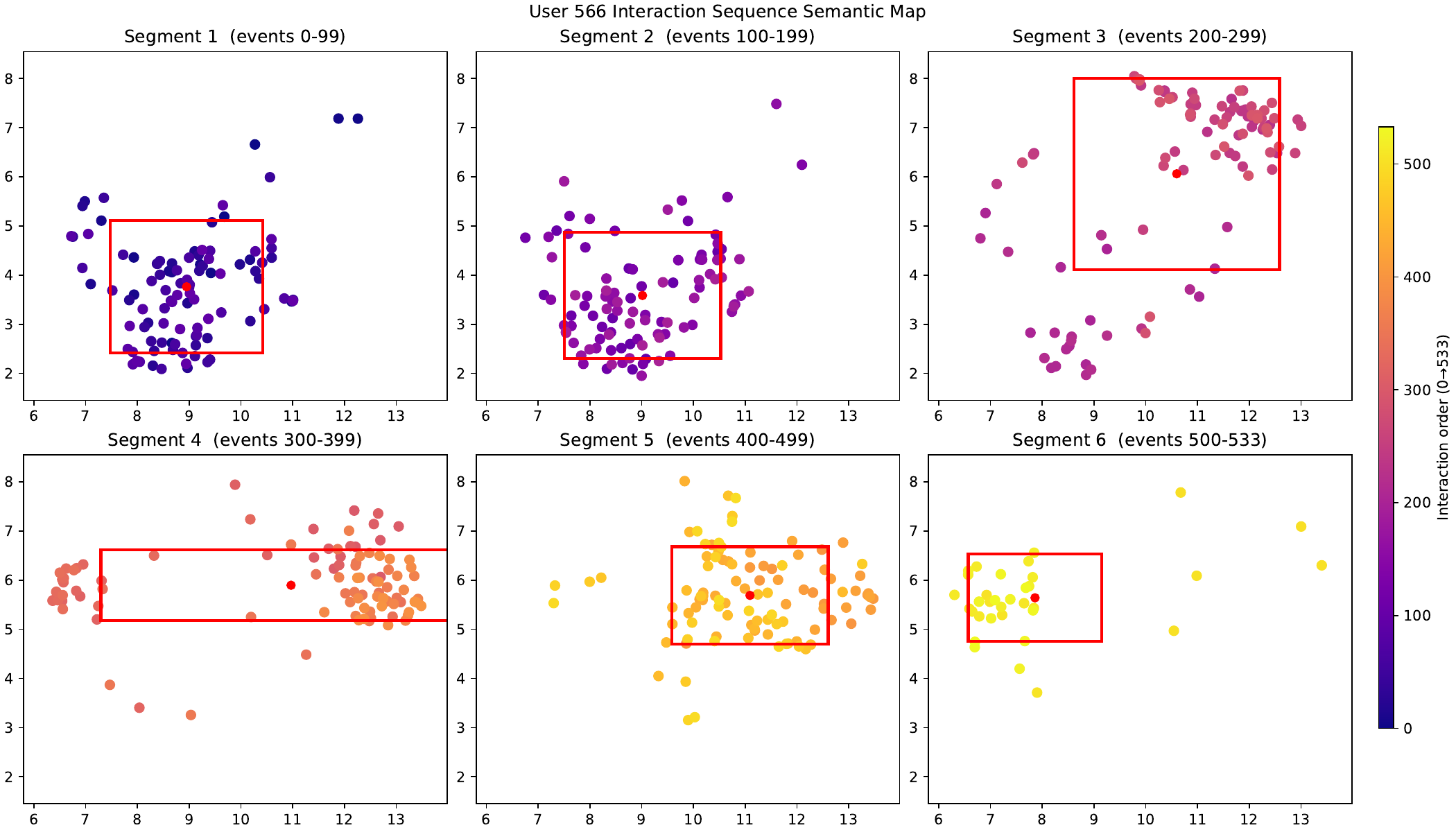}
  \caption{Case study of user 566 interaction sequence from ML-1M.}
  \label{fig:app1}
\end{figure*}
\label{sec:sparse}
To investigate whether interaction sequences can be processed in a block-wise pattern, we extracted the complete interaction sequence of user \#566 from ML-1M, which contains 534 sequential interactions. After partitioning the sequence into temporally contiguous blocks, we visualized the evolving representation of each block. The resulting maps (Figure~\ref{fig:app1}) reveal an apparent drift of user interests across the 534 events.  Moreover, for every block, we identified the cluster centroid and enclosed the 80\% of the interactions closest to that centroid (indicated by bounding boxes). The concentration of points within each box demonstrates that user interests remain relatively stable within specific temporal windows. This provides strong empirical evidence that the interaction sequence can be effectively divided into coherent blocks for subsequent modeling.

\section{Experiment Details}
\label{sec:efficienset}
To rapidly obtain efficiency results while still accommodating the longest possible sequences on the GPU for SASRec, we fixed the training and evaluation batch size at 32 and set the number of Transformer layers to one; all other hyperparameters remained unchanged during the efficiency experiments conducted across varying sequence lengths. Setting the sequence to 2000 under the inference of SASRec requires more memory, which our current hardware is not able to support (GPU 4090 with 23.64 GB memory).

\section{Inference and Optimization}
\label{sec:IO}
After obtaining item representations $\mathbf{H} \in \mathbb{R}^{L \times d}$ from the sequence through our Blossom Attention layers, we perform next-item recommendation by computing a probability distribution over the entire item vocabulary. At time step $t$, for each candidate item $v_i$, the recommendation score is calculated as:
\begin{equation}
    r_i = \mathbf{h}_t^T \mathbf{e}_{v_i}
\end{equation}

\noindent where $\mathbf{h}_t \in \mathbb{R}^d$ is the representation of the $t$-th position in the sequence (serving as the sequence representation), and $\mathbf{e}_{v_i} \in \mathbb{R}^d$ is the embedding of candidate item $v_i$. The predicted probability that the next item is $v_i$ is computed via softmax:

\begin{equation}
\hat{y}_i = \frac{\exp(r_i)}{\sum_{v_j \in \mathcal{V}} \exp(r_j)}
\end{equation}

\noindent where $\mathcal{V}$ denotes the item vocabulary. We formulate the sequential recommendation task as a cross-entropy optimization problem:
\begin{equation}
\mathcal{L}(y, \hat{y}) = -( y \log(\hat{y}) + (1-y)\log(1-\hat{y}))
\end{equation}

\noindent where $y \in \{0, 1\}$ is the ground truth label. The training objective is to learn optimal parameters for the network.

\section{\textbf{Efficiency Analysis}}
\label{sec:E Ana}
Beyond theoretical complexity, we analyze the actual computational efficiency from the perspective of participating interactions. 
In models such as SASRec~\cite{kang2018self} that adopt full attention, every interaction within a user's sequence attends the computation; herein, we show the theoretical number of participating interactions and the approximate reduction by using BlossomRec (Settings are as in Section~\ref{sec:expset}). Table~\ref{tab:token_participation} demonstrates that BlossomRec substantially curtails the volume of pairwise interactions relative to standard full-attention mechanisms, with the reduction becoming more pronounced as sequence length increases. 

\begin{table}[ht]
  \centering
  \caption{Analysis of participating interactions in attention computation.}
  \label{tab:token_participation}         
  \begin{tabular}{lcccc}
    \hline
    \textbf{Sequence Length} & \textbf{256} & \textbf{512} & \textbf{1024} & \textbf{2048} \\
    \hline
    \textbf{Full Attention \cite{vaswani2017attention}} & 256  & 512  & 1024 & 2048 \\
    \textbf{BlossomRec}                                & 103  & 120  & 153  & 218  \\
    \textbf{Reduction}                                   & 59.8\,\% & 76.6\,\% & 85.1\,\% & 89.4\,\% \\
    \hline
  \end{tabular}
\end{table}

\noindent This substantial reduction in interaction participation directly translates to faster inference, making BlossomRec practical for real-world applications with long user interaction histories.

\section{Implementation Details}
\label{sec:I Detail}
We set the embedding dimension to 128 for ML-1M, 64 for other datasets. The maximum sequence length is set to 200 for ML-1M, 100 for other datasets. We use Adam optimizer with a learning rate of 0.001, training and evaluation batch size of 2048, and 512 for Gowalla. Dropout rates are set to 0.2 for ML-1M, 0.3 for other datasets. For transformer-based models, layers are set to 2, heads are set to 8. The training process employs early stopping with a patience of 15 epochs based on NDCG@10. For BlossomRec-specific hyperparameters, we set compression size to 32, stride length to 16, selection block size to 16, window size to 8, and mask block size to 1. Other settings follow the default of recbole.

%% file: 0PaperID2079.bbl

\begin{thebibliography}{66}


\ifx \showCODEN    \undefined \def \showCODEN     #1{\unskip}     \fi
\ifx \showISBNx    \undefined \def \showISBNx     #1{\unskip}     \fi
\ifx \showISBNxiii \undefined \def \showISBNxiii  #1{\unskip}     \fi
\ifx \showISSN     \undefined \def \showISSN      #1{\unskip}     \fi
\ifx \showLCCN     \undefined \def \showLCCN      #1{\unskip}     \fi
\ifx \shownote     \undefined \def \shownote      #1{#1}          \fi
\ifx \showarticletitle \undefined \def \showarticletitle #1{#1}   \fi
\ifx \showURL      \undefined \def \showURL       {\relax}        \fi
\providecommand\bibfield[2]{#2}
\providecommand\bibinfo[2]{#2}
\providecommand\natexlab[1]{#1}
\providecommand\showeprint[2][]{arXiv:#2}

\bibitem[Ainslie et~al\mbox{.}(2023)]%
        {ainslie2023gqa}
\bibfield{author}{\bibinfo{person}{Joshua Ainslie}, \bibinfo{person}{James Lee-Thorp}, \bibinfo{person}{Michiel De~Jong}, \bibinfo{person}{Yury Zemlyanskiy}, \bibinfo{person}{Federico Lebr{\'o}n}, {and} \bibinfo{person}{Sumit Sanghai}.} \bibinfo{year}{2023}\natexlab{}.
\newblock \showarticletitle{Gqa: Training generalized multi-query transformer models from multi-head checkpoints}.
\newblock \bibinfo{journal}{\emph{arXiv preprint arXiv:2305.13245}} (\bibinfo{year}{2023}).
\newblock


\bibitem[Ba et~al\mbox{.}(2016)]%
        {ba2016layer}
\bibfield{author}{\bibinfo{person}{Jimmy~Lei Ba}, \bibinfo{person}{Jamie~Ryan Kiros}, {and} \bibinfo{person}{Geoffrey~E Hinton}.} \bibinfo{year}{2016}\natexlab{}.
\newblock \showarticletitle{Layer normalization}.
\newblock \bibinfo{journal}{\emph{arXiv preprint arXiv:1607.06450}} (\bibinfo{year}{2016}).
\newblock


\bibitem[Bao et~al\mbox{.}(2025)]%
        {bao2025bi}
\bibfield{author}{\bibinfo{person}{Keqin Bao}, \bibinfo{person}{Jizhi Zhang}, \bibinfo{person}{Wenjie Wang}, \bibinfo{person}{Yang Zhang}, \bibinfo{person}{Zhengyi Yang}, \bibinfo{person}{Yanchen Luo}, \bibinfo{person}{Chong Chen}, \bibinfo{person}{Fuli Feng}, {and} \bibinfo{person}{Qi Tian}.} \bibinfo{year}{2025}\natexlab{}.
\newblock \showarticletitle{A bi-step grounding paradigm for large language models in recommendation systems}.
\newblock \bibinfo{journal}{\emph{ACM Transactions on Recommender Systems}} \bibinfo{volume}{3}, \bibinfo{number}{4} (\bibinfo{year}{2025}), \bibinfo{pages}{1--27}.
\newblock


\bibitem[Beltagy et~al\mbox{.}(2020)]%
        {beltagy2020longformer}
\bibfield{author}{\bibinfo{person}{Iz Beltagy}, \bibinfo{person}{Matthew~E Peters}, {and} \bibinfo{person}{Arman Cohan}.} \bibinfo{year}{2020}\natexlab{}.
\newblock \showarticletitle{Longformer: The long-document transformer}.
\newblock \bibinfo{journal}{\emph{arXiv preprint arXiv:2004.05150}} (\bibinfo{year}{2020}).
\newblock


\bibitem[Chai et~al\mbox{.}(2025)]%
        {chai2025longer}
\bibfield{author}{\bibinfo{person}{Zheng Chai}, \bibinfo{person}{Qin Ren}, \bibinfo{person}{Xijun Xiao}, \bibinfo{person}{Huizhi Yang}, \bibinfo{person}{Bo Han}, \bibinfo{person}{Sijun Zhang}, \bibinfo{person}{Di Chen}, \bibinfo{person}{Hui Lu}, \bibinfo{person}{Wenlin Zhao}, \bibinfo{person}{Lele Yu}, {et~al\mbox{.}}} \bibinfo{year}{2025}\natexlab{}.
\newblock \showarticletitle{Longer: Scaling up long sequence modeling in industrial recommenders}. In \bibinfo{booktitle}{\emph{Proceedings of the Nineteenth ACM Conference on Recommender Systems}}. \bibinfo{pages}{247--256}.
\newblock


\bibitem[Chen et~al\mbox{.}(2024)]%
        {chen2024hllm}
\bibfield{author}{\bibinfo{person}{Junyi Chen}, \bibinfo{person}{Lu Chi}, \bibinfo{person}{Bingyue Peng}, {and} \bibinfo{person}{Zehuan Yuan}.} \bibinfo{year}{2024}\natexlab{}.
\newblock \showarticletitle{Hllm: Enhancing sequential recommendations via hierarchical large language models for item and user modeling}.
\newblock \bibinfo{journal}{\emph{arXiv preprint arXiv:2409.12740}} (\bibinfo{year}{2024}).
\newblock


\bibitem[Chen et~al\mbox{.}(2025)]%
        {chen2025powerattention}
\bibfield{author}{\bibinfo{person}{Lida Chen}, \bibinfo{person}{Dong Xu}, \bibinfo{person}{Chenxin An}, \bibinfo{person}{Xintao Wang}, \bibinfo{person}{Yikai Zhang}, \bibinfo{person}{Jiangjie Chen}, \bibinfo{person}{Zujie Liang}, \bibinfo{person}{Feng Wei}, \bibinfo{person}{Jiaqing Liang}, \bibinfo{person}{Yanghua Xiao}, {et~al\mbox{.}}} \bibinfo{year}{2025}\natexlab{}.
\newblock \showarticletitle{PowerAttention: Exponentially Scaling of Receptive Fields for Effective Sparse Attention}.
\newblock \bibinfo{journal}{\emph{arXiv preprint arXiv:2503.03588}} (\bibinfo{year}{2025}).
\newblock


\bibitem[Covington et~al\mbox{.}(2016)]%
        {covington2016youtube}
\bibfield{author}{\bibinfo{person}{Paul Covington}, \bibinfo{person}{Jay Adams}, {and} \bibinfo{person}{Emre Sargin}.} \bibinfo{year}{2016}\natexlab{}.
\newblock \showarticletitle{Deep neural networks for youtube recommendations}. In \bibinfo{booktitle}{\emph{Proceedings of the 10th ACM conference on recommender systems}}. \bibinfo{pages}{191--198}.
\newblock


\bibitem[Cui et~al\mbox{.}(2024)]%
        {cui2024distillation}
\bibfield{author}{\bibinfo{person}{Yu Cui}, \bibinfo{person}{Feng Liu}, \bibinfo{person}{Pengbo Wang}, \bibinfo{person}{Bohao Wang}, \bibinfo{person}{Heng Tang}, \bibinfo{person}{Yi Wan}, \bibinfo{person}{Jun Wang}, {and} \bibinfo{person}{Jiawei Chen}.} \bibinfo{year}{2024}\natexlab{}.
\newblock \showarticletitle{Distillation matters: empowering sequential recommenders to match the performance of large language models}. In \bibinfo{booktitle}{\emph{Proceedings of the 18th ACM Conference on Recommender Systems}}. \bibinfo{pages}{507--517}.
\newblock


\bibitem[Ding et~al\mbox{.}(2023)]%
        {ding2023longnet}
\bibfield{author}{\bibinfo{person}{Jiayu Ding}, \bibinfo{person}{Shuming Ma}, \bibinfo{person}{Li Dong}, \bibinfo{person}{Xingxing Zhang}, \bibinfo{person}{Shaohan Huang}, \bibinfo{person}{Wenhui Wang}, \bibinfo{person}{Nanning Zheng}, {and} \bibinfo{person}{Furu Wei}.} \bibinfo{year}{2023}\natexlab{}.
\newblock \showarticletitle{Longnet: Scaling transformers to 1,000,000,000 tokens}.
\newblock \bibinfo{journal}{\emph{arXiv preprint arXiv:2307.02486}} (\bibinfo{year}{2023}).
\newblock


\bibitem[Du et~al\mbox{.}(2022)]%
        {du2022contrastive}
\bibfield{author}{\bibinfo{person}{Hanwen Du}, \bibinfo{person}{Hui Shi}, \bibinfo{person}{Pengpeng Zhao}, \bibinfo{person}{Deqing Wang}, \bibinfo{person}{Victor~S Sheng}, \bibinfo{person}{Yanchi Liu}, \bibinfo{person}{Guanfeng Liu}, {and} \bibinfo{person}{Lei Zhao}.} \bibinfo{year}{2022}\natexlab{}.
\newblock \showarticletitle{Contrastive learning with bidirectional transformers for sequential recommendation}. In \bibinfo{booktitle}{\emph{Proceedings of the 31st ACM International Conference on Information \& Knowledge Management}}. \bibinfo{pages}{396--405}.
\newblock


\bibitem[Feng et~al\mbox{.}(2024)]%
        {feng2024long}
\bibfield{author}{\bibinfo{person}{Ningya Feng}, \bibinfo{person}{Junwei Pan}, \bibinfo{person}{Jialong Wu}, \bibinfo{person}{Baixu Chen}, \bibinfo{person}{Ximei Wang}, \bibinfo{person}{Qian Li}, \bibinfo{person}{Xian Hu}, \bibinfo{person}{Jie Jiang}, {and} \bibinfo{person}{Mingsheng Long}.} \bibinfo{year}{2024}\natexlab{}.
\newblock \showarticletitle{Long-Sequence Recommendation Models Need Decoupled Embeddings}.
\newblock \bibinfo{journal}{\emph{arXiv preprint arXiv:2410.02604}} (\bibinfo{year}{2024}).
\newblock


\bibitem[Fu et~al\mbox{.}(2025)]%
        {fu2025multimodal}
\bibfield{author}{\bibinfo{person}{Yongrui Fu}, \bibinfo{person}{Jian Liu}, \bibinfo{person}{Tao Li}, \bibinfo{person}{Zonggang Wu}, \bibinfo{person}{Shouke Qin}, {and} \bibinfo{person}{Hanmeng Liu}.} \bibinfo{year}{2025}\natexlab{}.
\newblock \showarticletitle{Multimodal Fusion And Sparse Attention-based Alignment Model for Long Sequential Recommendation}.
\newblock \bibinfo{journal}{\emph{arXiv preprint arXiv:2508.09664}} (\bibinfo{year}{2025}).
\newblock


\bibitem[Fu et~al\mbox{.}(2023)]%
        {fu2023unified}
\bibfield{author}{\bibinfo{person}{Zichuan Fu}, \bibinfo{person}{Xiangyang Li}, \bibinfo{person}{Chuhan Wu}, \bibinfo{person}{Yichao Wang}, \bibinfo{person}{Kuicai Dong}, \bibinfo{person}{Xiangyu Zhao}, \bibinfo{person}{Mengchen Zhao}, \bibinfo{person}{Huifeng Guo}, {and} \bibinfo{person}{Ruiming Tang}.} \bibinfo{year}{2023}\natexlab{}.
\newblock \showarticletitle{A unified framework for multi-domain ctr prediction via large language models}.
\newblock \bibinfo{journal}{\emph{ACM Transactions on Information Systems}} (\bibinfo{year}{2023}).
\newblock


\bibitem[Gao et~al\mbox{.}(2024a)]%
        {gao2024hierrec}
\bibfield{author}{\bibinfo{person}{Jingtong Gao}, \bibinfo{person}{Bo Chen}, \bibinfo{person}{Menghui Zhu}, \bibinfo{person}{Xiangyu Zhao}, \bibinfo{person}{Xiaopeng Li}, \bibinfo{person}{Yuhao Wang}, \bibinfo{person}{Yichao Wang}, \bibinfo{person}{Huifeng Guo}, {and} \bibinfo{person}{Ruiming Tang}.} \bibinfo{year}{2024}\natexlab{a}.
\newblock \showarticletitle{Hierrec: Scenario-aware hierarchical modeling for multi-scenario recommendations}. In \bibinfo{booktitle}{\emph{Proceedings of the 33rd ACM International Conference on Information and Knowledge Management}}. \bibinfo{pages}{653--662}.
\newblock


\bibitem[Gao et~al\mbox{.}(2025)]%
        {gao2025samplellm}
\bibfield{author}{\bibinfo{person}{Jingtong Gao}, \bibinfo{person}{Zhaocheng Du}, \bibinfo{person}{Xiaopeng Li}, \bibinfo{person}{Yichao Wang}, \bibinfo{person}{Xiangyang Li}, \bibinfo{person}{Huifeng Guo}, \bibinfo{person}{Ruiming Tang}, {and} \bibinfo{person}{Xiangyu Zhao}.} \bibinfo{year}{2025}\natexlab{}.
\newblock \showarticletitle{SampleLLM: Optimizing Tabular Data Synthesis in Recommendations}. In \bibinfo{booktitle}{\emph{Companion Proceedings of the ACM on Web Conference 2025}}. \bibinfo{pages}{211--220}.
\newblock


\bibitem[Gao et~al\mbox{.}(2024b)]%
        {gao2024smlp4rec}
\bibfield{author}{\bibinfo{person}{Jingtong Gao}, \bibinfo{person}{Xiangyu Zhao}, \bibinfo{person}{Muyang Li}, \bibinfo{person}{Minghao Zhao}, \bibinfo{person}{Runze Wu}, \bibinfo{person}{Ruocheng Guo}, \bibinfo{person}{Yiding Liu}, {and} \bibinfo{person}{Dawei Yin}.} \bibinfo{year}{2024}\natexlab{b}.
\newblock \showarticletitle{Smlp4rec: An efficient all-mlp architecture for sequential recommendations}.
\newblock \bibinfo{journal}{\emph{ACM Transactions on Information Systems}} \bibinfo{volume}{42}, \bibinfo{number}{3} (\bibinfo{year}{2024}), \bibinfo{pages}{1--23}.
\newblock


\bibitem[Geng et~al\mbox{.}(2024)]%
        {geng2024breaking}
\bibfield{author}{\bibinfo{person}{Binzong Geng}, \bibinfo{person}{Zhaoxin Huan}, \bibinfo{person}{Xiaolu Zhang}, \bibinfo{person}{Yong He}, \bibinfo{person}{Liang Zhang}, \bibinfo{person}{Fajie Yuan}, \bibinfo{person}{Jun Zhou}, {and} \bibinfo{person}{Linjian Mo}.} \bibinfo{year}{2024}\natexlab{}.
\newblock \showarticletitle{Breaking the length barrier: Llm-enhanced CTR prediction in long textual user behaviors}. In \bibinfo{booktitle}{\emph{Proceedings of the 47th International ACM SIGIR Conference on Research and Development in Information Retrieval}}. \bibinfo{pages}{2311--2315}.
\newblock


\bibitem[Gu and Dao(2023)]%
        {gu2023mamba}
\bibfield{author}{\bibinfo{person}{Albert Gu} {and} \bibinfo{person}{Tri Dao}.} \bibinfo{year}{2023}\natexlab{}.
\newblock \showarticletitle{Mamba: Linear-time sequence modeling with selective state spaces}.
\newblock \bibinfo{journal}{\emph{arXiv preprint arXiv:2312.00752}} (\bibinfo{year}{2023}).
\newblock


\bibitem[Guo et~al\mbox{.}(2019)]%
        {guo2019star}
\bibfield{author}{\bibinfo{person}{Qipeng Guo}, \bibinfo{person}{Xipeng Qiu}, \bibinfo{person}{Pengfei Liu}, \bibinfo{person}{Yunfan Shao}, \bibinfo{person}{Xiangyang Xue}, {and} \bibinfo{person}{Zheng Zhang}.} \bibinfo{year}{2019}\natexlab{}.
\newblock \showarticletitle{Star-transformer}.
\newblock \bibinfo{journal}{\emph{arXiv preprint arXiv:1902.09113}} (\bibinfo{year}{2019}).
\newblock


\bibitem[Huang et~al\mbox{.}(2018)]%
        {huang2018csan}
\bibfield{author}{\bibinfo{person}{Xiaowen Huang}, \bibinfo{person}{Shengsheng Qian}, \bibinfo{person}{Quan Fang}, \bibinfo{person}{Jitao Sang}, {and} \bibinfo{person}{Changsheng Xu}.} \bibinfo{year}{2018}\natexlab{}.
\newblock \showarticletitle{Csan: Contextual self-attention network for user sequential recommendation}. In \bibinfo{booktitle}{\emph{Proceedings of the 26th ACM international conference on Multimedia}}. \bibinfo{pages}{447--455}.
\newblock


\bibitem[Jannach and Ludewig(2017)]%
        {jannach2017recurrent}
\bibfield{author}{\bibinfo{person}{Dietmar Jannach} {and} \bibinfo{person}{Malte Ludewig}.} \bibinfo{year}{2017}\natexlab{}.
\newblock \showarticletitle{When recurrent neural networks meet the neighborhood for session-based recommendation}. In \bibinfo{booktitle}{\emph{Proceedings of the eleventh ACM conference on recommender systems}}. \bibinfo{pages}{306--310}.
\newblock


\bibitem[Jia et~al\mbox{.}(2024)]%
        {jia2024d3}
\bibfield{author}{\bibinfo{person}{Pengyue Jia}, \bibinfo{person}{Yichao Wang}, \bibinfo{person}{Shanru Lin}, \bibinfo{person}{Xiaopeng Li}, \bibinfo{person}{Xiangyu Zhao}, \bibinfo{person}{Huifeng Guo}, {and} \bibinfo{person}{Ruiming Tang}.} \bibinfo{year}{2024}\natexlab{}.
\newblock \showarticletitle{D3: A methodological exploration of domain division, modeling, and balance in multi-domain recommendations}. In \bibinfo{booktitle}{\emph{Proceedings of the AAAI Conference on Artificial Intelligence}}, Vol.~\bibinfo{volume}{38}. \bibinfo{pages}{8553--8561}.
\newblock


\bibitem[Jiang et~al\mbox{.}(2023)]%
        {SWA}
\bibfield{author}{\bibinfo{person}{Albert~Q. Jiang}, \bibinfo{person}{Alexandre Sablayrolles}, \bibinfo{person}{Arthur Mensch}, \bibinfo{person}{Chris Bamford}, \bibinfo{person}{Devendra~Singh Chaplot}, \bibinfo{person}{de~las~Diego Casas}, \bibinfo{person}{Florian Bressand}, \bibinfo{person}{Gianna Lengyel}, \bibinfo{person}{Guillaume Lample}, \bibinfo{person}{Lucile Saulnier}, \bibinfo{person}{Lélio~Renard Lavaud}, \bibinfo{person}{Marie-Anne Lachaux}, \bibinfo{person}{Pierre Stock}, \bibinfo{person}{Teven~Le Scao}, \bibinfo{person}{Thibaut Lavril}, \bibinfo{person}{Thomas Wang}, \bibinfo{person}{Timothée Lacroix}, {and} \bibinfo{person}{William~El Sayed}.} \bibinfo{year}{2023}\natexlab{}.
\newblock \showarticletitle{Mistral 7B}.
\newblock \bibinfo{journal}{\emph{arxiv:2310.06825 [cs.CL,cs.AI,cs.LG]}} (\bibinfo{date}{10 10} \bibinfo{year}{2023}).
\newblock
\newblock
\shownote{[Online; accessed 2025-10-08]}.


\bibitem[Jiang et~al\mbox{.}(2024)]%
        {jiang2024minference}
\bibfield{author}{\bibinfo{person}{Huiqiang Jiang}, \bibinfo{person}{Yucheng Li}, \bibinfo{person}{Chengruidong Zhang}, \bibinfo{person}{Qianhui Wu}, \bibinfo{person}{Xufang Luo}, \bibinfo{person}{Surin Ahn}, \bibinfo{person}{Zhenhua Han}, \bibinfo{person}{Amir~H Abdi}, \bibinfo{person}{Dongsheng Li}, \bibinfo{person}{Chin-Yew Lin}, {et~al\mbox{.}}} \bibinfo{year}{2024}\natexlab{}.
\newblock \showarticletitle{Minference 1.0: Accelerating pre-filling for long-context llms via dynamic sparse attention}.
\newblock \bibinfo{journal}{\emph{Advances in Neural Information Processing Systems}}  \bibinfo{volume}{37} (\bibinfo{year}{2024}), \bibinfo{pages}{52481--52515}.
\newblock


\bibitem[Kang and McAuley(2018)]%
        {kang2018self}
\bibfield{author}{\bibinfo{person}{Wang-Cheng Kang} {and} \bibinfo{person}{Julian McAuley}.} \bibinfo{year}{2018}\natexlab{}.
\newblock \showarticletitle{Self-attentive sequential recommendation}. In \bibinfo{booktitle}{\emph{2018 IEEE international conference on data mining (ICDM)}}. IEEE, \bibinfo{pages}{197--206}.
\newblock


\bibitem[Li et~al\mbox{.}(2023b)]%
        {li2023strec}
\bibfield{author}{\bibinfo{person}{Chengxi Li}, \bibinfo{person}{Yejing Wang}, \bibinfo{person}{Qidong Liu}, \bibinfo{person}{Xiangyu Zhao}, \bibinfo{person}{Wanyu Wang}, \bibinfo{person}{Yiqi Wang}, \bibinfo{person}{Lixin Zou}, \bibinfo{person}{Wenqi Fan}, {and} \bibinfo{person}{Qing Li}.} \bibinfo{year}{2023}\natexlab{b}.
\newblock \showarticletitle{STRec: Sparse transformer for sequential recommendations}. In \bibinfo{booktitle}{\emph{Proceedings of the 17th ACM conference on recommender systems}}. \bibinfo{pages}{101--111}.
\newblock


\bibitem[Li et~al\mbox{.}(2026)]%
        {li2026collectivekv}
\bibfield{author}{\bibinfo{person}{Jingyu Li}, \bibinfo{person}{Zhaocheng Du}, \bibinfo{person}{Qianhui Zhu}, \bibinfo{person}{Zhicheng Zhang}, \bibinfo{person}{Song-Li Wu}, \bibinfo{person}{Chaolang Li}, \bibinfo{person}{Pengwen Dai}, {et~al\mbox{.}}} \bibinfo{year}{2026}\natexlab{}.
\newblock \showarticletitle{CollectiveKV: Decoupling and Sharing Collaborative Information in Sequential Recommendation}.
\newblock \bibinfo{journal}{\emph{arXiv preprint arXiv:2601.19178}} (\bibinfo{year}{2026}).
\newblock


\bibitem[Li et~al\mbox{.}(2020)]%
        {li2020time}
\bibfield{author}{\bibinfo{person}{Jiacheng Li}, \bibinfo{person}{Yujie Wang}, {and} \bibinfo{person}{Julian McAuley}.} \bibinfo{year}{2020}\natexlab{}.
\newblock \showarticletitle{Time interval aware self-attention for sequential recommendation}. In \bibinfo{booktitle}{\emph{Proceedings of the 13th international conference on web search and data mining}}. \bibinfo{pages}{322--330}.
\newblock


\bibitem[Li et~al\mbox{.}(2023d)]%
        {li2023automlp}
\bibfield{author}{\bibinfo{person}{Muyang Li}, \bibinfo{person}{Zijian Zhang}, \bibinfo{person}{Xiangyu Zhao}, \bibinfo{person}{Wanyu Wang}, \bibinfo{person}{Minghao Zhao}, \bibinfo{person}{Runze Wu}, {and} \bibinfo{person}{Ruocheng Guo}.} \bibinfo{year}{2023}\natexlab{d}.
\newblock \showarticletitle{Automlp: Automated mlp for sequential recommendations}. In \bibinfo{booktitle}{\emph{Proceedings of the ACM web conference 2023}}. \bibinfo{pages}{1190--1198}.
\newblock


\bibitem[Li et~al\mbox{.}(2022b)]%
        {li2022mlp4rec}
\bibfield{author}{\bibinfo{person}{Muyang Li}, \bibinfo{person}{Xiangyu Zhao}, \bibinfo{person}{Chuan Lyu}, \bibinfo{person}{Minghao Zhao}, \bibinfo{person}{Runze Wu}, {and} \bibinfo{person}{Ruocheng Guo}.} \bibinfo{year}{2022}\natexlab{b}.
\newblock \showarticletitle{MLP4Rec: A pure MLP architecture for sequential recommendations}.
\newblock \bibinfo{journal}{\emph{arXiv preprint arXiv:2204.11510}} (\bibinfo{year}{2022}).
\newblock


\bibitem[Li et~al\mbox{.}(2019)]%
        {li2019logsparse}
\bibfield{author}{\bibinfo{person}{Shiyang Li}, \bibinfo{person}{Xiaoyong Jin}, \bibinfo{person}{Yao Xuan}, \bibinfo{person}{Xiyou Zhou}, \bibinfo{person}{Wenhu Chen}, \bibinfo{person}{Yu-Xiang Wang}, {and} \bibinfo{person}{Xifeng Yan}.} \bibinfo{year}{2019}\natexlab{}.
\newblock \showarticletitle{Enhancing the locality and breaking the memory bottleneck of transformer on time series forecasting}.
\newblock \bibinfo{journal}{\emph{Advances in neural information processing systems}}  \bibinfo{volume}{32} (\bibinfo{year}{2019}).
\newblock


\bibitem[Li et~al\mbox{.}(2025)]%
        {li2025survey}
\bibfield{author}{\bibinfo{person}{Xiaopeng Li}, \bibinfo{person}{Bo Chen}, \bibinfo{person}{Junda She}, \bibinfo{person}{Shiteng Cao}, \bibinfo{person}{You Wang}, \bibinfo{person}{Qinlin Jia}, \bibinfo{person}{Haiying He}, \bibinfo{person}{Zheli Zhou}, \bibinfo{person}{Zhao Liu}, \bibinfo{person}{Ji Liu}, {et~al\mbox{.}}} \bibinfo{year}{2025}\natexlab{}.
\newblock \showarticletitle{A Survey of Generative Recommendation from a Tri-Decoupled Perspective: Tokenization, Architecture, and Optimization}.
\newblock  (\bibinfo{year}{2025}).
\newblock


\bibitem[Li et~al\mbox{.}(2022a)]%
        {li2022gromov}
\bibfield{author}{\bibinfo{person}{Xinhang Li}, \bibinfo{person}{Zhaopeng Qiu}, \bibinfo{person}{Xiangyu Zhao}, \bibinfo{person}{Zihao Wang}, \bibinfo{person}{Yong Zhang}, \bibinfo{person}{Chunxiao Xing}, {and} \bibinfo{person}{Xian Wu}.} \bibinfo{year}{2022}\natexlab{a}.
\newblock \showarticletitle{Gromov-wasserstein guided representation learning for cross-domain recommendation}. In \bibinfo{booktitle}{\emph{Proceedings of the 31st ACM International Conference on Information \& Knowledge Management}}. \bibinfo{pages}{1199--1208}.
\newblock


\bibitem[Li et~al\mbox{.}(2023a)]%
        {li2023agent4ranking}
\bibfield{author}{\bibinfo{person}{Xiaopeng Li}, \bibinfo{person}{Lixin Su}, \bibinfo{person}{Pengyue Jia}, \bibinfo{person}{Xiangyu Zhao}, \bibinfo{person}{Suqi Cheng}, \bibinfo{person}{Junfeng Wang}, {and} \bibinfo{person}{Dawei Yin}.} \bibinfo{year}{2023}\natexlab{a}.
\newblock \showarticletitle{Agent4ranking: Semantic robust ranking via personalized query rewriting using multi-agent llm}.
\newblock \bibinfo{journal}{\emph{arXiv preprint arXiv:2312.15450}} (\bibinfo{year}{2023}).
\newblock


\bibitem[Li et~al\mbox{.}(2023c)]%
        {li2023hamur}
\bibfield{author}{\bibinfo{person}{Xiaopeng Li}, \bibinfo{person}{Fan Yan}, \bibinfo{person}{Xiangyu Zhao}, \bibinfo{person}{Yichao Wang}, \bibinfo{person}{Bo Chen}, \bibinfo{person}{Huifeng Guo}, {and} \bibinfo{person}{Ruiming Tang}.} \bibinfo{year}{2023}\natexlab{c}.
\newblock \showarticletitle{Hamur: Hyper adapter for multi-domain recommendation}. In \bibinfo{booktitle}{\emph{Proceedings of the 32nd ACM International Conference on Information and Knowledge Management}}. \bibinfo{pages}{1268--1277}.
\newblock


\bibitem[Liu et~al\mbox{.}(2024a)]%
        {liu2024deepseek}
\bibfield{author}{\bibinfo{person}{Aixin Liu}, \bibinfo{person}{Bei Feng}, \bibinfo{person}{Bing Xue}, \bibinfo{person}{Bingxuan Wang}, \bibinfo{person}{Bochao Wu}, \bibinfo{person}{Chengda Lu}, \bibinfo{person}{Chenggang Zhao}, \bibinfo{person}{Chengqi Deng}, \bibinfo{person}{Chenyu Zhang}, \bibinfo{person}{Chong Ruan}, {et~al\mbox{.}}} \bibinfo{year}{2024}\natexlab{a}.
\newblock \showarticletitle{Deepseek-v3 technical report}.
\newblock \bibinfo{journal}{\emph{arXiv preprint arXiv:2412.19437}} (\bibinfo{year}{2024}).
\newblock


\bibitem[Liu et~al\mbox{.}(2024b)]%
        {liu2024behavior}
\bibfield{author}{\bibinfo{person}{Chengkai Liu}, \bibinfo{person}{Jianghao Lin}, \bibinfo{person}{Hanzhou Liu}, \bibinfo{person}{Jianling Wang}, {and} \bibinfo{person}{James Caverlee}.} \bibinfo{year}{2024}\natexlab{b}.
\newblock \showarticletitle{Behavior-dependent linear recurrent units for efficient sequential recommendation}. In \bibinfo{booktitle}{\emph{Proceedings of the 33rd ACM international conference on information and knowledge management}}. \bibinfo{pages}{1430--1440}.
\newblock


\bibitem[Liu et~al\mbox{.}(2024c)]%
        {liu2024mamba4rec}
\bibfield{author}{\bibinfo{person}{Chengkai Liu}, \bibinfo{person}{Jianghao Lin}, \bibinfo{person}{Jianling Wang}, \bibinfo{person}{Hanzhou Liu}, {and} \bibinfo{person}{James Caverlee}.} \bibinfo{year}{2024}\natexlab{c}.
\newblock \showarticletitle{Mamba4rec: Towards efficient sequential recommendation with selective state space models}.
\newblock \bibinfo{journal}{\emph{arXiv preprint arXiv:2403.03900}} (\bibinfo{year}{2024}).
\newblock


\bibitem[Liu et~al\mbox{.}(2023b)]%
        {liu2023linrec}
\bibfield{author}{\bibinfo{person}{Langming Liu}, \bibinfo{person}{Liu Cai}, \bibinfo{person}{Chi Zhang}, \bibinfo{person}{Xiangyu Zhao}, \bibinfo{person}{Jingtong Gao}, \bibinfo{person}{Wanyu Wang}, \bibinfo{person}{Yifu Lv}, \bibinfo{person}{Wenqi Fan}, \bibinfo{person}{Yiqi Wang}, \bibinfo{person}{Ming He}, {et~al\mbox{.}}} \bibinfo{year}{2023}\natexlab{b}.
\newblock \showarticletitle{Linrec: Linear attention mechanism for long-term sequential recommender systems}. In \bibinfo{booktitle}{\emph{Proceedings of the 46th International ACM SIGIR Conference on Research and Development in Information Retrieval}}. \bibinfo{pages}{289--299}.
\newblock


\bibitem[Liu et~al\mbox{.}(2024d)]%
        {liu2024llm}
\bibfield{author}{\bibinfo{person}{Qidong Liu}, \bibinfo{person}{Xian Wu}, \bibinfo{person}{Yejing Wang}, \bibinfo{person}{Zijian Zhang}, \bibinfo{person}{Feng Tian}, \bibinfo{person}{Yefeng Zheng}, {and} \bibinfo{person}{Xiangyu Zhao}.} \bibinfo{year}{2024}\natexlab{d}.
\newblock \showarticletitle{Llm-esr: Large language models enhancement for long-tailed sequential recommendation}.
\newblock \bibinfo{journal}{\emph{Advances in Neural Information Processing Systems}}  \bibinfo{volume}{37} (\bibinfo{year}{2024}), \bibinfo{pages}{26701--26727}.
\newblock


\bibitem[Liu et~al\mbox{.}(2025b)]%
        {liu2025bridge}
\bibfield{author}{\bibinfo{person}{Qidong Liu}, \bibinfo{person}{Xiangyu Zhao}, \bibinfo{person}{Yejing Wang}, \bibinfo{person}{Zijian Zhang}, \bibinfo{person}{Howard Zhong}, \bibinfo{person}{Chong Chen}, \bibinfo{person}{Xiang Li}, \bibinfo{person}{Wei Huang}, {and} \bibinfo{person}{Feng Tian}.} \bibinfo{year}{2025}\natexlab{b}.
\newblock \showarticletitle{Bridge the Domains: Large Language Models Enhanced Cross-domain Sequential Recommendation}. In \bibinfo{booktitle}{\emph{Proceedings of the 48th International ACM SIGIR Conference on Research and Development in Information Retrieval}}. \bibinfo{pages}{1582--1592}.
\newblock


\bibitem[Liu et~al\mbox{.}(2023a)]%
        {liu2023exploration}
\bibfield{author}{\bibinfo{person}{Shuchang Liu}, \bibinfo{person}{Qingpeng Cai}, \bibinfo{person}{Bowen Sun}, \bibinfo{person}{Yuhao Wang}, \bibinfo{person}{Ji Jiang}, \bibinfo{person}{Dong Zheng}, \bibinfo{person}{Peng Jiang}, \bibinfo{person}{Kun Gai}, \bibinfo{person}{Xiangyu Zhao}, {and} \bibinfo{person}{Yongfeng Zhang}.} \bibinfo{year}{2023}\natexlab{a}.
\newblock \showarticletitle{Exploration and regularization of the latent action space in recommendation}. In \bibinfo{booktitle}{\emph{Proceedings of the ACM Web Conference 2023}}. \bibinfo{pages}{833--844}.
\newblock


\bibitem[Liu et~al\mbox{.}(2025a)]%
        {liu2025sigma}
\bibfield{author}{\bibinfo{person}{Ziwei Liu}, \bibinfo{person}{Qidong Liu}, \bibinfo{person}{Yejing Wang}, \bibinfo{person}{Wanyu Wang}, \bibinfo{person}{Pengyue Jia}, \bibinfo{person}{Maolin Wang}, \bibinfo{person}{Zitao Liu}, \bibinfo{person}{Yi Chang}, {and} \bibinfo{person}{Xiangyu Zhao}.} \bibinfo{year}{2025}\natexlab{a}.
\newblock \showarticletitle{SIGMA: Selective Gated Mamba for Sequential Recommendation}. In \bibinfo{booktitle}{\emph{Proceedings of the AAAI Conference on Artificial Intelligence}}, Vol.~\bibinfo{volume}{39}. \bibinfo{pages}{12264--12272}.
\newblock


\bibitem[Lu et~al\mbox{.}(2025b)]%
        {lu2025moba}
\bibfield{author}{\bibinfo{person}{Enzhe Lu}, \bibinfo{person}{Zhejun Jiang}, \bibinfo{person}{Jingyuan Liu}, \bibinfo{person}{Yulun Du}, \bibinfo{person}{Tao Jiang}, \bibinfo{person}{Chao Hong}, \bibinfo{person}{Shaowei Liu}, \bibinfo{person}{Weiran He}, \bibinfo{person}{Enming Yuan}, \bibinfo{person}{Yuzhi Wang}, {et~al\mbox{.}}} \bibinfo{year}{2025}\natexlab{b}.
\newblock \showarticletitle{Moba: Mixture of block attention for long-context llms}.
\newblock \bibinfo{journal}{\emph{arXiv preprint arXiv:2502.13189}} (\bibinfo{year}{2025}).
\newblock


\bibitem[Lu et~al\mbox{.}(2025a)]%
        {lu2025liveforesighter}
\bibfield{author}{\bibinfo{person}{Yucheng Lu}, \bibinfo{person}{Jiangxia Cao}, \bibinfo{person}{Xu Kuan}, \bibinfo{person}{Wei Cheng}, \bibinfo{person}{Wei Jiang}, \bibinfo{person}{Jiaming Zhang}, \bibinfo{person}{Yang Shuang}, \bibinfo{person}{Liu Zhaojie}, {and} \bibinfo{person}{Liyin Hong}.} \bibinfo{year}{2025}\natexlab{a}.
\newblock \showarticletitle{LiveForesighter: Generating Future Information for Live-Streaming Recommendations at Kuaishou}.
\newblock \bibinfo{journal}{\emph{arXiv preprint arXiv:2502.06557}} (\bibinfo{year}{2025}).
\newblock


\bibitem[Lv et~al\mbox{.}(2019)]%
        {lv2019sdm}
\bibfield{author}{\bibinfo{person}{Fuyu Lv}, \bibinfo{person}{Taiwei Jin}, \bibinfo{person}{Changlong Yu}, \bibinfo{person}{Fei Sun}, \bibinfo{person}{Quan Lin}, \bibinfo{person}{Keping Yang}, {and} \bibinfo{person}{Wilfred Ng}.} \bibinfo{year}{2019}\natexlab{}.
\newblock \showarticletitle{SDM: Sequential deep matching model for online large-scale recommender system}. In \bibinfo{booktitle}{\emph{Proceedings of the 28th ACM international conference on information and knowledge management}}. \bibinfo{pages}{2635--2643}.
\newblock


\bibitem[Ma et~al\mbox{.}(2024)]%
        {ma2024block}
\bibfield{author}{\bibinfo{person}{Dongyang Ma}, \bibinfo{person}{Yan Wang}, {and} \bibinfo{person}{Lan Tian}.} \bibinfo{year}{2024}\natexlab{}.
\newblock \showarticletitle{Block-attention for efficient prefilling}.
\newblock \bibinfo{journal}{\emph{arXiv preprint arXiv:2409.15355}} (\bibinfo{year}{2024}).
\newblock


\bibitem[Shen et~al\mbox{.}(2022)]%
        {shen2022hierarchically}
\bibfield{author}{\bibinfo{person}{Qijie Shen}, \bibinfo{person}{Hong Wen}, \bibinfo{person}{Jing Zhang}, {and} \bibinfo{person}{Qi Rao}.} \bibinfo{year}{2022}\natexlab{}.
\newblock \showarticletitle{Hierarchically fusing long and short-term user interests for click-through rate prediction in product search}. In \bibinfo{booktitle}{\emph{Proceedings of the 31st ACM International Conference on Information \& Knowledge Management}}. \bibinfo{pages}{1767--1776}.
\newblock


\bibitem[Song et~al\mbox{.}(2025)]%
        {song2025videonsa}
\bibfield{author}{\bibinfo{person}{Enxin Song}, \bibinfo{person}{Wenhao Chai}, \bibinfo{person}{Shusheng Yang}, \bibinfo{person}{Ethan Armand}, \bibinfo{person}{Xiaojun Shan}, \bibinfo{person}{Haiyang Xu}, \bibinfo{person}{Jianwen Xie}, {and} \bibinfo{person}{Zhuowen Tu}.} \bibinfo{year}{2025}\natexlab{}.
\newblock \showarticletitle{Videonsa: Native sparse attention scales video understanding}.
\newblock \bibinfo{journal}{\emph{arXiv preprint arXiv:2510.02295}} (\bibinfo{year}{2025}).
\newblock


\bibitem[Su et~al\mbox{.}(2024)]%
        {su2024roformer}
\bibfield{author}{\bibinfo{person}{Jianlin Su}, \bibinfo{person}{Murtadha Ahmed}, \bibinfo{person}{Yu Lu}, \bibinfo{person}{Shengfeng Pan}, \bibinfo{person}{Wen Bo}, {and} \bibinfo{person}{Yunfeng Liu}.} \bibinfo{year}{2024}\natexlab{}.
\newblock \showarticletitle{Roformer: Enhanced transformer with rotary position embedding}.
\newblock \bibinfo{journal}{\emph{Neurocomputing}}  \bibinfo{volume}{568} (\bibinfo{year}{2024}), \bibinfo{pages}{127063}.
\newblock


\bibitem[Sun et~al\mbox{.}(2019)]%
        {sun2019bert4rec}
\bibfield{author}{\bibinfo{person}{Fei Sun}, \bibinfo{person}{Jun Liu}, \bibinfo{person}{Jian Wu}, \bibinfo{person}{Changhua Pei}, \bibinfo{person}{Xiao Lin}, \bibinfo{person}{Wenwu Ou}, {and} \bibinfo{person}{Peng Jiang}.} \bibinfo{year}{2019}\natexlab{}.
\newblock \showarticletitle{BERT4Rec: Sequential recommendation with bidirectional encoder representations from transformer}. In \bibinfo{booktitle}{\emph{Proceedings of the 28th ACM international conference on information and knowledge management}}. \bibinfo{pages}{1441--1450}.
\newblock


\bibitem[Tillet et~al\mbox{.}(2019)]%
        {tillet2019triton}
\bibfield{author}{\bibinfo{person}{Philippe Tillet}, \bibinfo{person}{Hsiang-Tsung Kung}, {and} \bibinfo{person}{David Cox}.} \bibinfo{year}{2019}\natexlab{}.
\newblock \showarticletitle{Triton: an intermediate language and compiler for tiled neural network computations}. In \bibinfo{booktitle}{\emph{Proceedings of the 3rd ACM SIGPLAN International Workshop on Machine Learning and Programming Languages}}. \bibinfo{pages}{10--19}.
\newblock


\bibitem[Vaswani et~al\mbox{.}(2017)]%
        {vaswani2017attention}
\bibfield{author}{\bibinfo{person}{Ashish Vaswani}, \bibinfo{person}{Noam Shazeer}, \bibinfo{person}{Niki Parmar}, \bibinfo{person}{Jakob Uszkoreit}, \bibinfo{person}{Llion Jones}, \bibinfo{person}{Aidan~N Gomez}, \bibinfo{person}{{\L}ukasz Kaiser}, {and} \bibinfo{person}{Illia Polosukhin}.} \bibinfo{year}{2017}\natexlab{}.
\newblock \showarticletitle{Attention is all you need}.
\newblock \bibinfo{journal}{\emph{Advances in neural information processing systems}}  \bibinfo{volume}{30} (\bibinfo{year}{2017}).
\newblock


\bibitem[Wang et~al\mbox{.}(2023)]%
        {wang2023plate}
\bibfield{author}{\bibinfo{person}{Yuhao Wang}, \bibinfo{person}{Xiangyu Zhao}, \bibinfo{person}{Bo Chen}, \bibinfo{person}{Qidong Liu}, \bibinfo{person}{Huifeng Guo}, \bibinfo{person}{Huanshuo Liu}, \bibinfo{person}{Yichao Wang}, \bibinfo{person}{Rui Zhang}, {and} \bibinfo{person}{Ruiming Tang}.} \bibinfo{year}{2023}\natexlab{}.
\newblock \showarticletitle{PLATE: A prompt-enhanced paradigm for multi-scenario recommendations}. In \bibinfo{booktitle}{\emph{Proceedings of the 46th International ACM SIGIR Conference on Research and Development in Information Retrieval}}. \bibinfo{pages}{1498--1507}.
\newblock


\bibitem[Yu et~al\mbox{.}(2026)]%
        {yu2026malloc}
\bibfield{author}{\bibinfo{person}{Qihang Yu}, \bibinfo{person}{Kairui Fu}, \bibinfo{person}{Zhaocheng Du}, \bibinfo{person}{Yuxuan Si}, \bibinfo{person}{Kaiyuan Li}, \bibinfo{person}{Weihao Zhao}, \bibinfo{person}{Zhicheng Zhang}, \bibinfo{person}{Jieming Zhu}, \bibinfo{person}{Quanyu Dai}, \bibinfo{person}{Zhenhua Dong}, {et~al\mbox{.}}} \bibinfo{year}{2026}\natexlab{}.
\newblock \showarticletitle{MALLOC: Benchmarking the Memory-aware Long Sequence Compression for Large Sequential Recommendation}.
\newblock \bibinfo{journal}{\emph{arXiv preprint arXiv:2601.20234}} (\bibinfo{year}{2026}).
\newblock


\bibitem[Yuan et~al\mbox{.}(2025)]%
        {yuan2025native}
\bibfield{author}{\bibinfo{person}{Jingyang Yuan}, \bibinfo{person}{Huazuo Gao}, \bibinfo{person}{Damai Dai}, \bibinfo{person}{Junyu Luo}, \bibinfo{person}{Liang Zhao}, \bibinfo{person}{Zhengyan Zhang}, \bibinfo{person}{Zhenda Xie}, \bibinfo{person}{YX Wei}, \bibinfo{person}{Lean Wang}, \bibinfo{person}{Zhiping Xiao}, {et~al\mbox{.}}} \bibinfo{year}{2025}\natexlab{}.
\newblock \showarticletitle{Native sparse attention: Hardware-aligned and natively trainable sparse attention}.
\newblock \bibinfo{journal}{\emph{arXiv preprint arXiv:2502.11089}} (\bibinfo{year}{2025}).
\newblock


\bibitem[Zaheer et~al\mbox{.}(2020)]%
        {zaheer2020big}
\bibfield{author}{\bibinfo{person}{Manzil Zaheer}, \bibinfo{person}{Guru Guruganesh}, \bibinfo{person}{Kumar~Avinava Dubey}, \bibinfo{person}{Joshua Ainslie}, \bibinfo{person}{Chris Alberti}, \bibinfo{person}{Santiago Ontanon}, \bibinfo{person}{Philip Pham}, \bibinfo{person}{Anirudh Ravula}, \bibinfo{person}{Qifan Wang}, \bibinfo{person}{Li Yang}, {et~al\mbox{.}}} \bibinfo{year}{2020}\natexlab{}.
\newblock \showarticletitle{Big bird: Transformers for longer sequences}.
\newblock \bibinfo{journal}{\emph{Advances in neural information processing systems}}  \bibinfo{volume}{33} (\bibinfo{year}{2020}), \bibinfo{pages}{17283--17297}.
\newblock


\bibitem[Zhai et~al\mbox{.}(2024)]%
        {zhai2024actions}
\bibfield{author}{\bibinfo{person}{Jiaqi Zhai}, \bibinfo{person}{Lucy Liao}, \bibinfo{person}{Xing Liu}, \bibinfo{person}{Yueming Wang}, \bibinfo{person}{Rui Li}, \bibinfo{person}{Xuan Cao}, \bibinfo{person}{Leon Gao}, \bibinfo{person}{Zhaojie Gong}, \bibinfo{person}{Fangda Gu}, \bibinfo{person}{Michael He}, {et~al\mbox{.}}} \bibinfo{year}{2024}\natexlab{}.
\newblock \showarticletitle{Actions speak louder than words: Trillion-parameter sequential transducers for generative recommendations}.
\newblock \bibinfo{journal}{\emph{arXiv preprint arXiv:2402.17152}} (\bibinfo{year}{2024}).
\newblock


\bibitem[Zhang et~al\mbox{.}(2024)]%
        {zhang2024notellm}
\bibfield{author}{\bibinfo{person}{Chao Zhang}, \bibinfo{person}{Haoxin Zhang}, \bibinfo{person}{Shiwei Wu}, \bibinfo{person}{Di Wu}, \bibinfo{person}{Tong Xu}, \bibinfo{person}{Xiangyu Zhao}, \bibinfo{person}{Yan Gao}, \bibinfo{person}{Yao Hu}, {and} \bibinfo{person}{Enhong Chen}.} \bibinfo{year}{2024}\natexlab{}.
\newblock \showarticletitle{Notellm-2: Multimodal large representation models for recommendation}.
\newblock \bibinfo{journal}{\emph{arXiv preprint arXiv:2405.16789}} (\bibinfo{year}{2024}).
\newblock


\bibitem[Zhang et~al\mbox{.}(2025a)]%
        {zhang2025m2rec}
\bibfield{author}{\bibinfo{person}{Qianru Zhang}, \bibinfo{person}{Liang Qu}, \bibinfo{person}{Honggang Wen}, \bibinfo{person}{Dong Huang}, \bibinfo{person}{Siu-Ming Yiu}, \bibinfo{person}{Nguyen Quoc~Viet Hung}, {and} \bibinfo{person}{Hongzhi Yin}.} \bibinfo{year}{2025}\natexlab{a}.
\newblock \showarticletitle{M2Rec: Multi-scale Mamba for Efficient Sequential Recommendation}.
\newblock \bibinfo{journal}{\emph{arXiv preprint arXiv:2505.04445}} (\bibinfo{year}{2025}).
\newblock


\bibitem[Zhang et~al\mbox{.}(2025b)]%
        {zhang2025glint}
\bibfield{author}{\bibinfo{person}{Sheng Zhang}, \bibinfo{person}{Maolin Wang}, \bibinfo{person}{Wanyu Wang}, \bibinfo{person}{Jingtong Gao}, \bibinfo{person}{Xiangyu Zhao}, \bibinfo{person}{Yu Yang}, \bibinfo{person}{Xuetao Wei}, \bibinfo{person}{Zitao Liu}, {and} \bibinfo{person}{Tong Xu}.} \bibinfo{year}{2025}\natexlab{b}.
\newblock \showarticletitle{Glint-ru: Gated lightweight intelligent recurrent units for sequential recommender systems}. In \bibinfo{booktitle}{\emph{Proceedings of the 31st ACM SIGKDD Conference on Knowledge Discovery and Data Mining V. 1}}. \bibinfo{pages}{1948--1959}.
\newblock


\bibitem[Zhang et~al\mbox{.}(2026)]%
        {zhang2026length}
\bibfield{author}{\bibinfo{person}{Zhicheng Zhang}, \bibinfo{person}{Zhaocheng Du}, \bibinfo{person}{Jieming Zhu}, \bibinfo{person}{Jiwei Tang}, \bibinfo{person}{Fengyuan Lu}, \bibinfo{person}{Wang Jiaheng}, \bibinfo{person}{Song-Li Wu}, \bibinfo{person}{Qianhui Zhu}, \bibinfo{person}{Jingyu Li}, \bibinfo{person}{Hai-Tao Zheng}, {et~al\mbox{.}}} \bibinfo{year}{2026}\natexlab{}.
\newblock \showarticletitle{Length-Adaptive Interest Network for Balancing Long and Short Sequence Modeling in CTR Prediction}.
\newblock \bibinfo{journal}{\emph{arXiv preprint arXiv:2601.19142}} (\bibinfo{year}{2026}).
\newblock


\bibitem[Zhao et~al\mbox{.}(2025)]%
        {zhao2025joint}
\bibfield{author}{\bibinfo{person}{Xiangyu Zhao}, \bibinfo{person}{Yichao Wang}, \bibinfo{person}{Bo Chen}, \bibinfo{person}{Jingtong Gao}, \bibinfo{person}{Yuhao Wang}, \bibinfo{person}{Xiaopeng Li}, \bibinfo{person}{Pengyue Jia}, \bibinfo{person}{Qidong Liu}, \bibinfo{person}{Huifeng Guo}, {and} \bibinfo{person}{Ruiming Tang}.} \bibinfo{year}{2025}\natexlab{}.
\newblock \showarticletitle{Joint Modeling in Recommendations: A Survey}.
\newblock \bibinfo{journal}{\emph{arXiv preprint arXiv:2502.21195}} (\bibinfo{year}{2025}).
\newblock


\bibitem[Zhao et~al\mbox{.}(2018a)]%
        {zhao2018deep}
\bibfield{author}{\bibinfo{person}{Xiangyu Zhao}, \bibinfo{person}{Long Xia}, \bibinfo{person}{Liang Zhang}, \bibinfo{person}{Zhuoye Ding}, \bibinfo{person}{Dawei Yin}, {and} \bibinfo{person}{Jiliang Tang}.} \bibinfo{year}{2018}\natexlab{a}.
\newblock \showarticletitle{Deep reinforcement learning for page-wise recommendations}. In \bibinfo{booktitle}{\emph{Proceedings of the 12th ACM conference on recommender systems}}. \bibinfo{pages}{95--103}.
\newblock


\bibitem[Zhao et~al\mbox{.}(2018b)]%
        {zhao2018recommendations}
\bibfield{author}{\bibinfo{person}{Xiangyu Zhao}, \bibinfo{person}{Liang Zhang}, \bibinfo{person}{Zhuoye Ding}, \bibinfo{person}{Long Xia}, \bibinfo{person}{Jiliang Tang}, {and} \bibinfo{person}{Dawei Yin}.} \bibinfo{year}{2018}\natexlab{b}.
\newblock \showarticletitle{Recommendations with negative feedback via pairwise deep reinforcement learning}. In \bibinfo{booktitle}{\emph{Proceedings of the 24th ACM SIGKDD international conference on knowledge discovery \& data mining}}. \bibinfo{pages}{1040--1048}.
\newblock


\end{thebibliography}
